\documentclass[a4paper,fleqn]{cas-sc}

\usepackage[numbers]{natbib}
\usepackage{xcolor}
\usepackage{amsmath}
\usepackage{palatino}
\usepackage{subfig}
\usepackage{amsfonts}
\usepackage{graphicx}
\usepackage{mathtools}
\usepackage{booktabs}
\usepackage[justification=centering,font=scriptsize]{caption}
\usepackage{microtype}
\usepackage[ruled,vlined,]{algorithm2e}
\usepackage{lineno}

\def\tsc#1{\csdef{#1}{\textsc{\lowercase{#1}}\xspace}}
\tsc{WGM}
\tsc{QE}
\tsc{EP}
\tsc{PMS}
\tsc{BEC}
\tsc{DE}
\tsc{IXPE}

\begin{document}
\let\WriteBookmarks\relax
\def\floatpagepagefraction{1}
\def\textpagefraction{.001}
\shorttitle{DL for X-ray Polarimetry}
\shortauthors{A.L.Peirson et~al.}

\title [mode = title]{Deep Ensemble Analysis for Imaging X-ray Polarimetry}



\author[1]{A.L.Peirson}[type=editor,
                     orcid=0000-0001-6292-1911]
\cormark[1]
\ead{alpv95@stanford.edu}


\address[1]{Department of Physics \& Kavli Institute for Particle Astrophysics and Cosmology, Stanford, CA, 94305}

\author[1]{R.W.Romani}

\author[2]{H.L.Marshall}[%
   ]


\address[2]{Kavli Institute for Astrophysics and Space Research, MIT, 77 Massachusetts Ave., Cambridge, MA, 02139.
}

\author%
[3]
{J.F.Steiner}

\address[3]{Harvard-Smithsonian Center for Astrophysics, 60 Garden Street, Cambridge, MA 02138, USA.}
    
\author%
[4]
{L.Baldini}

\address[4]{Universit\'a di Pisa and INFN-Sezione di Pisa, Pisa, Italy, I-56127}

\cortext[cor1]{Corresponding author}

\begin{abstract}
We present a method for enhancing the sensitivity of X-ray telescopic observations with imaging polarimeters, with a focus on the gas pixel detectors (GPDs) to be flown on the Imaging X-ray Polarimetry Explorer (IXPE). Our analysis determines photoelectron directions, X-ray absorption points and X-ray energies for 1-9\,keV event tracks, with estimates for both the statistical and model (reconstruction) uncertainties. We use a weighted maximum likelihood combination of predictions from a deep ensemble of ResNet convolutional neural networks, trained on Monte Carlo event simulations. We define a figure of merit to compare the polarization bias-variance trade-off in track reconstruction algorithms. For power-law source spectra, our method improves on the current planned IXPE analysis (and previous deep learning approaches), providing $\sim45$\% increase in effective exposure times. For individual energies, our method produces 20-30\% absolute improvements in modulation factor for simulated 100\% polarized events, while keeping residual systematic modulation within $1\sigma$ of the finite sample minimum. Absorption point location and photon energy estimates are also significantly improved. We have validated our method with sample data from real GPD detectors. 

\end{abstract}




\maketitle

\section{Introduction}

X-ray polarization measurements offer rich opportunities to probe the magnetic field topology and emission physics of high energy astrophysical sources \citep{krawczynski_using_2019}. However, in the classical soft X-ray band (1-10 keV) such measurements have long been elusive, with only the Crab nebula providing a solid detection \citep{weisskopf_measurement_1976}. Happily, the recent development of photo-electron tracking detectors \citep{costa_efficient_2001} has greatly improved soft X-ray polarimetry prospects. These imaging X-ray polarimeters offer lower background and better control of systematic signals, and allow the study of important extended sources, such as Supernova Remnants (SNR) and Pulsar Wind Nebulae (PWNe). The gas pixel detector (GPD) \citep[see][]{bellazzini_sealed_2007} has brought this capability to the PolarLight CubeSat test \citep{feng_x-ray_2020}, the scheduled IXPE mission \citep{sgro_imaging_2019}, and the potential Chinese mission, eXTP \citep{zhang_extp_2017}.

IXPE \citep[][planned for launch in 2021]{weisskopf_overview_2018,odell_imaging_2018} will use three co-aligned X-ray telescopes, whose focal planes are imaged by GPDs with hexagonal pixels. IXPE's sensitivity is limited by the track analysis algorithm used to recover source polarization, spatial structure and energy, given a measured set of electron track images. In this work, we demonstrate a substantial improvement over the current state of-the-art track reconstruction. While the results shown here are specific to IXPE's GPDs, the methods are general, and can be applied to other imaging detector geometries. 

In the $1-10$ keV range, the cross-section for photoelectron emission is proportional to cos$^2(\theta)$, where $\theta$ is the angle between the normal incidence X-ray's electric vector position angle (EVPA) and the azimuthal emission direction of the photoelectron. By measuring a large number of individual photoelectron emission angles, one can recover the above distribution to extract the source polarization parameters: polarization fraction ($0\% \leq \Pi \leq 100\%$, or equivalently $0 \leq \Pi \leq 1$) and EVPA ($-\pi/2 \leq \phi < \pi/2$). In practice, the recovery of photoelectron emission angles from track images is imperfect. Track images are noisy due to Coulomb scattering and diffusion, and, especially for low energies, are often barely resolved from the Bragg peak emission at their ends. In many cases a secondary Auger electron track further complicates the analysis. The modulation factor $\mu_{100}$, defined as the recovered $\Pi$ for a 100\% polarized source, provides a useful description of the quality of the image reconstruction. For a measured source the true polarization is then given by calibrating: $\Pi_{\rm true} = \Pi_{\rm meas} / \mu_{100}$. In practice, 
track reconstruction challenges make $\mu_{100}$ highly energy dependent \citep{muleri_spectral_2010}.

Critically, track reconstruction methods must not introduce significant bias for an unpolarized source ($\Pi = 0$). For IXPE the hexagonal GPD pixels and a $120^\circ$ rotation between the telescopes are designed to minimize such x-y systematic biases. Thus the efficacy of an X-ray polarimeter depends on the recovered $\mu$ at both $100\%$ and $0\%$ polarization.
For imaging X-ray polarimeters one also wishes to reconstruct the X-ray absorption (conversion) points and event energy.
\begin{figure}[]
\centering
\includegraphics[width=0.57\textwidth]{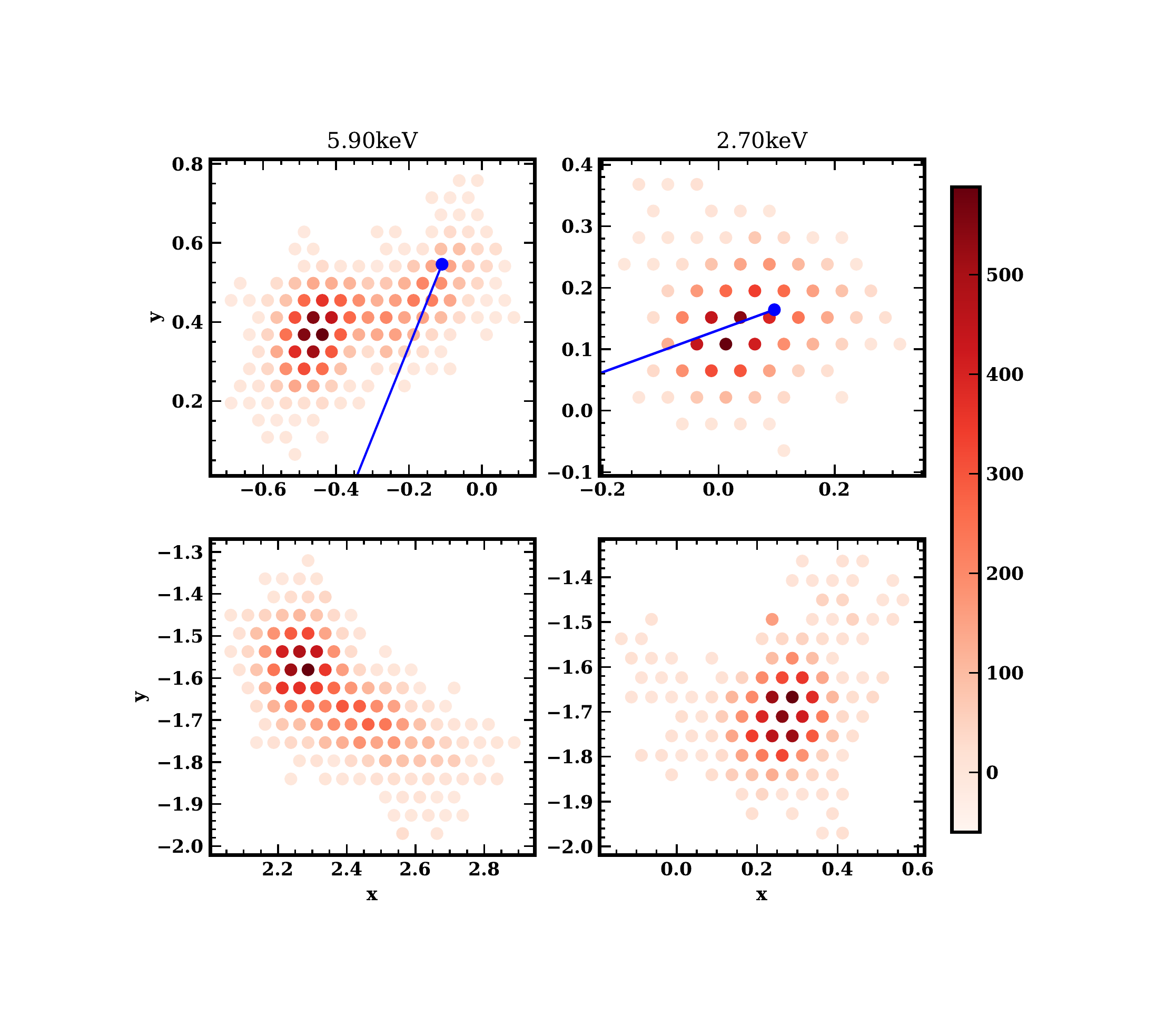}
\caption{Example IXPE GPD electron track images. The colormap represents charge deposited in each pixel. The top row is simulated (5.9\,keV, 2.7\,keV), the bottom shows real events of similar energy. For the simulated events, the blue dots and lines mark the true absorption point and initial $e^-$ direction. Lower energy tracks are typically smaller and less elliptical, thus harder to reconstruct.}
\label{fig:hex}
\end{figure} 

The current track reconstruction method for the GPD is a moment analysis described by \citet{bellazzini_novel_2003}. Impressive accuracies for the absorption point and EVPA angle are achieved from a simple re-weighted combination of track moments, with the track barycenter replacing the moment localization at low energies. Track energy estimates are proportional to the total collected GPD charge. The track ellipticity also provides a rough predictor for track reconstruction quality. High ellipticity tracks typically have more accurate angle estimates. However, simple moments cannot capture all of the image information, especially for long high energy tracks, and so a more sophisticated image analysis scheme should allow improved track parameters, as well as better assessment of reconstruction quality. Recently, machine learning techniques have been discussed as useful for X-ray polarization measurements \citep{moriakov_inferring_2020}; this is an ideal problem for such image analysis. 

\subsection{Deep learning for X-ray polarimetry}

Deep neural networks (NNs) have achieved state-of-the-art performance on a wide variety of machine learning tasks and are becoming increasingly popular in domains such as speech recognition \citep{graves_connectionist_2006}, natural language processing \citep{young_recent_2018}, bioinformatics \citep{tang_recent_2019} and especially computer vision \citep{krizhevsky_imagenet_2012}. Going from track images to numerical estimates can be classified as a computer vision problem, so it is not surprising that NNs would be well suited to track reconstruction. The Cherenkov Telescope Array (CTA) \cite{brill_investigating_2019} team have applied related deep learning methods to differentiate between cosmic rays and gamma rays. Notably they also have to deal with a hexagonal pixel grid. The IceCube collaboration has begun use of graph neural networks to identify 3D neutrino tracks with great success \citep{choma_graph_2018}. 

\citet{kitaguchi_convolutional_2019} have recently described a NN photoelectron track analysis for the non-imaging detector geometry that was intended for the PRAXyS X-ray polarimetry mission \citep{tamagawa_x-ray_2017}. Using convolutional neural networks (CNNs), they show significant improvements in modulation factor over a standard moment analysis for a square pixel grid polarimeter while maintaining $\lesssim 1\%$ modulation for unpolarized data. While an excellent start, this analysis did not recover energies, showed unexplained biases at the $1\%$ level, used event cuts rather than weights and provide only a binned polarization analysis. They also trained for only a handful of event energies and were not able to validate against real detector data. We have been able to improve on this analysis addressing all of the issues above while delivering superior performance across a wide (and continuous) range of energies.

The cornerstone of our approach involves deep ensembles \citep{lakshminarayanan_simple_2017}. These not only provide more accurate and less biased estimates than single NNs, but also give state-of-the-art estimates of predictive uncertainty. Using uncertainties in each of our track angle estimates, we developed an unbinned weighted maximum likelihood estimate (WMLE) approach to determine the final polarization parameters. This removes the need for excising data, making use of all measurements. With bootstrap analysis \citep{efron_introduction_1994} we are able to infer the final error on our polarization estimates and define an appropriate figure-of-merit (FoM) to compare different track reconstruction approaches.

This paper describes our track reconstruction algorithm using deep ensembles. Section 2 explains the extraction of angles, absorption points and energies from individual tracks. Section 3 briefly outlines our NN training procedure and selection.
In section 4 we describe our WMLE approach that takes an ensemble of track angles and uncertainties to final polarization parameters and their errors. There we define a figure-of-merit to compare the moment analysis, our approach and that of \citet{kitaguchi_convolutional_2019}. Section 5 shows our results and their interpretation. We conclude the study, mentioning prospects for additional improvements, in section 6. 

\section{Deep ensembles for track reconstruction}

We considered end-to-end deep learning approaches (as suggested by \citet{kitaguchi_convolutional_2019}) to go directly from a set of tracks to source polarization. There are a number of difficulties with this approach. Training would be very expensive (perhaps infeasible) and it is not obvious how to account for different track ensemble sizes. Also the observer may wish to adjust partitioning of the data set into e.g. time, energy and spatial subsets, with differing polarization. The best partition will often not be obvious before analysis starts, thus combining the properties of individual tracks allows a more efficient exploration of binning options. Finally detector-dependent artifacts can best be handled from individual tracks (with fine spatial positioning), rather than point spread function (PSF) weighted ensembles. Forming such ensembles in a second analysis tier allows additional flexibility.

For these reasons we do not consider a direct end-to-end approach but use a two step process: (1) extract features from individual tracks (angles, uncertainties, absorption points, energies) (2) combine an ensemble of the features to measure the final polarization statistics. However we do use the expected properties of polarized and unpolarized ensembles to guide the training and select the most effective networks.  We first describe step (1): the event characterization.

\begin{figure*}
\centering
\includegraphics[width=1.05\textwidth]{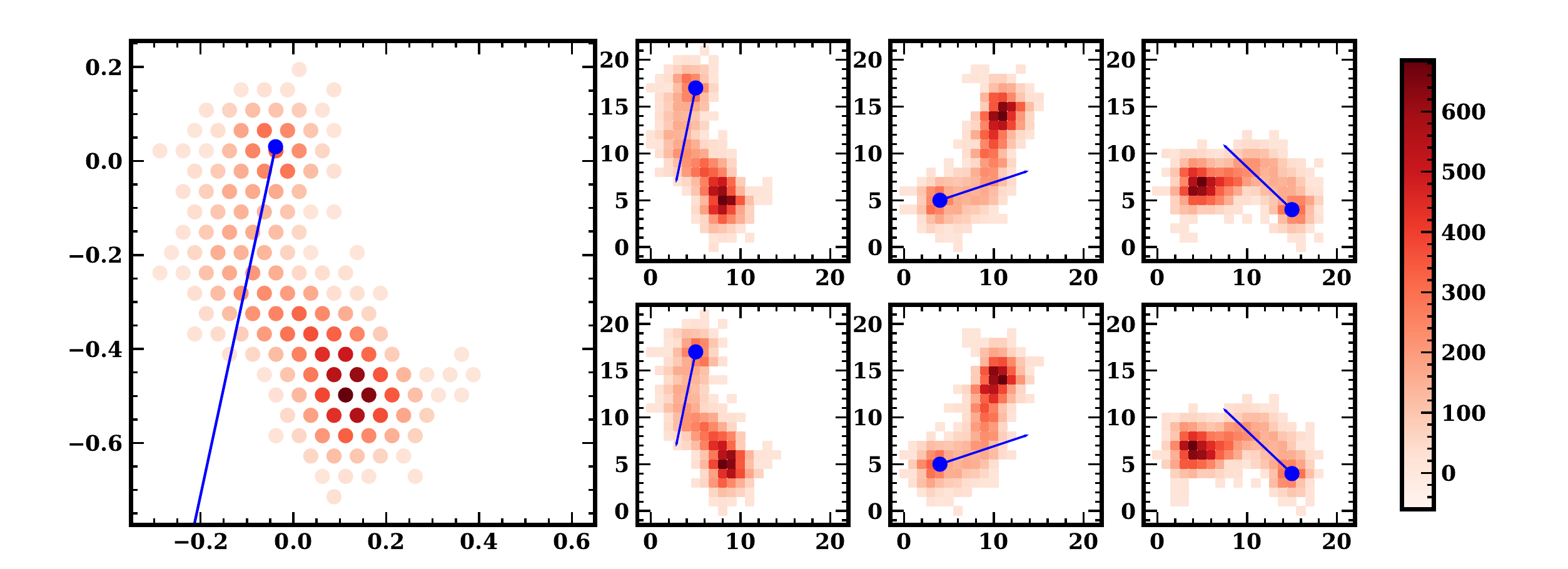}
\captionsetup{justification=centering,skip=2pt,font=scriptsize}
\caption{Example square conversions of a 6.4 keV hexagonal track (left panel). The six panels to the right show shifts along the 120$^\circ$ GPD axes; shifting odd rows (upper) or even rows (lower). For each hexagonal track, NNs are fed column-wise pairs of square conversions, along with the energy, absorption point (blue dot) and initial photoelectron direction (blue line) as labels.}
\label{fig:square}
\end{figure*}

\subsection{Deep Ensembles}

To extract the angles, absorption points and energies from individual tracks we use a supervised deep learning technique known as deep ensembles \citep{lakshminarayanan_simple_2017}. Deep ensembles are made up of an ensemble of individual NNs, each trained independently on the same data set to predict the desired output features. It has been shown that different random initializations of the same NN at the start of training leads to widely different prediction functions \citep{fort_deep_2019}. Deep ensembles exploit this property by incorporating the results of many differently initialized NNs, increasing the diversity of predictors. Considering all of the individual NN predictions together leads to a more robust, accurate,  and generalizable model with better uncertainty estimates \citep{ovadia_can_2019}.

In our case, we have an image to feature(s) regression problem. CNNs have been designed with an inductive bias appropriate for image regression problems. So our deep ensemble will be made up of individual CNNs. 

Deep ensembles provide estimates of the predictive uncertainty. There are two germane types of uncertainty one can model \citep{kendall_what_2017}. `Aleatoric uncertainty' captures noise inherent in the observations. This is equivalent to statistical uncertainty. On the other hand, `epistemic uncertainty' accounts for uncertainty in the model parameters – uncertainty which captures our ignorance about which model generated our collected data. This uncertainty can be reduced given enough data, and is often referred to as model uncertainty. We will model both of these uncertainties using deep ensembles and use them in our final polarization predictions in \textsection4.

\subsection{Hexagonal to square conversion}

The hexagonal grid used in the IXPE GPDs is designed to minimize polarization systematics, since hexapolar grid effects are orthogonal to the quadrupolar polarization signal. Example imaged photoelectron tracks at different energies are shown in fig.~\ref{fig:hex}.
Unfortunately, a hexagonal grid is not natively compatible with typical CNNs. It is possible to transform from a hexagonal to a square grid, however a naive transformation can lead to polarization biases and suboptimal NN performance. This is partly because the CNN convolutional kernels are not spatially equivariant in hexagonal space.

There are two main ways of converting between hexagonal and square grids: interpolation and pixel shifting. Interpolation places a fine square grid on top of the hexagonal image and interpolates. We avoid using interpolation since it adds noise to the raw data and is not easily reversible. Pixel shifting rearranges pixels by shifting alternate rows and then rescaling. The HexagDLy \citep{steppa_hexagdly_2019} software, designed for use in CTA, allows standard CNNs to operate on hexagonal images. It does this by pixel shifting the images to square arrays and then applying specialized convolutional kernels that preserve equivariance in hexagonal space. Unfortunately, in practice, this method proved too slow to train for the large event sets required for polarization estimation. Accordingly, we simply pixel shift each track along each of the six hexagonal axes (to avoid bias). Hexagonal tracks are rotated so that rows align horizontally (this can be done in three different ways, separated by $120^{\circ}$), then alternate rows are shifted (this can be done in two different ways, left and right) so that the track resembles a rectangular grid, as in HexagDLy. We convert the rectangular grid into a square image by defining the leftmost track pixel and bottom track pixel as the left edge and the base of the image respectively. We use a square image size of 50x50 pixels to fit all track sizes for energies up to 9 keV.
Since square track images are defined independently of the absolute hexagonal coordinate values, the initial hexagonal track rotation can be performed about any axis.

A single hexagonal track produces six square conversions (fig.~\ref{fig:square}), two for each $120^{\circ}$ angle.
A single training example for the NNs is formed by stacking the corresponding square conversion pair, similarly to color channels in a \textit{rgb} image CNN problem -- in this case with only two channels. At test time all 3 pairs are evaluated by the NNs and the predicted angles are rotated back to their original direction. 




It should be noted that clean track images, such as those shown in fig.~\ref{fig:hex}, already require a set of thresholding and clustering steps to isolate individual tracks from a detector (or simulation) snapshot. These are at present handled by IXPE's GPD software. The track rectification into the six square projections is simply an additional pre-processing step that we apply in collecting a dataset for supervised training.

\subsection{Training criteria}

In a typical CNN regression problem, during training the CNN takes as input a single image $\mathbf{x}_i$ with feature label $y_i$ and outputs single prediction $\hat{y}_i$. The NN parameters are optimized to minimize to the mean squared error (MSE) on the training data set $\Sigma^{N}_{i=1}(y_i - \hat{y}(\mathbf{x}_i))^2 / N$. In order to model the statistical uncertainty in predictions, individual networks in a deep ensemble each minimize the negative log-likelihood:
\begin{equation}
\label{eqn:DE_loss}
    L(y_i\mid\mathbf{x}_i) = \frac{{\rm log}(\hat{\sigma}^2(\mathbf{x}_i))}{2} + \frac{\|y_i - \hat{y}(\mathbf{x}_i)\|_2^2}{2\hat{\sigma}^2(\mathbf{x}_i)}.
\end{equation}
where $\hat{y}(\mathbf{x}_i)$ corresponds to the predicted mean and $\hat{\sigma}(\mathbf{x}_i)^2$ to the predicted variance. The $L_2$-norm is denoted by $\|.\|_2$.
Thus the NN produces an estimate of the feature and its statistical error $(\hat{y}(\mathbf{x}_i),\hat{\sigma}(\mathbf{x}_i))$. In practice however we train the NN to predict the log variance $\hat{s}(\mathbf{x}_i) = {\rm log}(\hat{\sigma}^2(\mathbf{x}_i))$
\begin{equation}
\label{eqn:log_loss}
    L(y_i\mid\mathbf{x}_i) = \frac{\hat{s}(\mathbf{x}_i)}{2} + \frac{e^{-\hat{s}(\mathbf{x}_i)}\|y_i - \hat{y}(\mathbf{x}_i)\|_2^2}{2}
\end{equation}
since this is more numerically stable \citep{kendall_what_2017}. 

We estimate the track angle and its statistical error with the following loss function,
\begin{equation}
\label{eqn:DE_phi_loss}
    L_{\theta}(\theta_i\mid\mathbf{x}_i) = \frac{\hat{s}(\mathbf{x}_i)}{2} + \frac{e^{-\hat{s}(\mathbf{x}_i)}}{2}\|\mathbf{v}_2^i - \mathbf{\hat{v}}_2(\mathbf{x}_i)\|_2^2 + \alpha.\|\mathbf{v}_1^i - \mathbf{\hat{v}}_1(\mathbf{x}_i)\|_2^2
\end{equation}
where $\mathbf{v}_2^i = ({\rm cos}2\theta_i,{\rm sin}2\theta_i)$ and $\mathbf{v}_1^i = ({\rm cos}\theta_i,{\rm sin}\theta_i)$ play the roles of the features $y_i$.
We parameterize the track angle $\theta_i$ as a unit 2D vector $\mathbf{v}_1^i$ to incorporate periodicity. 
By including a dipolar loss term  $\|\mathbf{v}_2^i - \mathbf{\hat{v}}_2 (\mathbf{x}_i)\|_2^2$, we account for the $180^{\circ}$ EVPA ambiguity. Without this additional term the sign of the principal axis for poorly resolved tracks is ambiguous and the CNN hedges its bets by selecting $\hat{\theta}$ orthogonal to the principal axis; low energy polarization resolution suffers. The hyperparameter $\alpha$ controls the relative importance of the monopole and dipole loss terms. Since polarization estimation depends only on $2\theta_i$, we estimate the statistical error only for the dipole term. \citet{kitaguchi_convolutional_2019} include only the dipole term in their loss function. Recovering the full directional information is, however, important for reducing detector bias and aids in post-processing analysis; a loss function with both terms produces the best results.

Since we also want to measure the absorption point and event energy, we include the two additional loss function terms:
\begin{equation}
     \label{eqn:DE_abs_loss}
    L_{\rm abs}(x_i,y_i\mid\mathbf{x}_i) = \frac{1}{2}\|(x_i,y_i) - (\hat{x}(\mathbf{x}_i),\hat{y}(\mathbf{x}_i))\|_2^2
\end{equation}
\begin{equation}
 \label{eqn:DE_E_loss}
L_{\rm E}(E_i\mid\mathbf{x}_i) =
    \begin{cases}
    \frac{1}{2}\big(E_i - \hat{E}(\mathbf{x}_i)\big)^2,& \text{if } E_i - \hat{E}(\mathbf{x}_i)\leq \epsilon\\
    \epsilon|E_i - \hat{E}(\mathbf{x}_i)| - \frac{1}{2}\epsilon^2,              & \text{otherwise}
\end{cases}
\end{equation}

where $E_i$ are the event energies and $x_i,y_i$ the coordinates of the absorption point in the rectified grid. We use an asymmetric Huber loss function (eq.~\ref{eqn:DE_E_loss}) for the energy to avoid high energy tails in predictions. 
These are destructive to IXPE's energy resolution since astrophysical spectra and IXPE's effective area will yield significantly more low energy events. The same result can be effected by training the NNs on a power law of track energies, instead of a flat distribution. The hyperparameter $\epsilon$ controls the degree of asymmetry.
Since our primary objective is polarization, we do not include separate location and energy statistical error parameters in the minimization. Nevertheless, as we will show, this method demonstrably improves localization and energy resolution.

In sum, our total multi-component loss function is
\begin{equation}
\label{eqn:DE_tot_loss}
    L(\theta_i,x_i,y_i,E_i\mid\mathbf{x}_i) = L_{\theta} + \beta L_{\rm abs}+ \gamma L_{\rm E}(\epsilon) + \delta \|w\|_2
\end{equation}
with hyperparameters $\alpha, \beta, \gamma, \delta, \epsilon$ that are tuned during training (\textsection4). The final term is a $L_2$-norm regularization on the NN parameters $w$. This is a common machine learning regularization scheme which prevents overfitting. Each individual CNN is trained to minimize eq.~\ref{eqn:DE_tot_loss}. The CNN computations take individual (square) track images $\mathbf{x}_i$ as input and output the feature vector \begin{equation}
\label{eqn:features}
({\rm cos}\theta_i,{\rm sin}\theta_i, \sigma^{\mathbf{s}}_i, x_i, y_i, E_i),
\end{equation}
where superscript $\mathbf{s}$ in $\sigma^{\mathbf{s}}_i$ denotes the predicted statistical, or aleatoric, error.

In practice, after training a set of NNs, we select a best-performing subset for inclusion in the final deep ensemble. This selection is described in \textsection3. We can model the epistemic uncertainty in our feature estimates by combining the final feature predictions from all the NNs in our ensemble. For each track, $\mathbf{x_i}$, a deep ensemble of $M$ networks $j = 1, \ldots, M$ produces $M$ output feature vectors (eq.~\ref{eqn:features}).
We calculate the epistemic uncertainty as the standard deviation of $2\theta_{ij}$ over the NN ensemble predictions $j$. Since $2\theta_{ij}$ are periodic, we use a von Mises distribution to estimate the standard deviation in $2\theta_{ij}$ over $j$:
\begin{equation}
    R_i = \left[ \frac{1}{M}\sum\limits_{j=1}^M {\rm cos}2\theta_{ij},
    \frac{1}{M}\sum\limits_{j=1}^M {\rm sin}2\theta_{ij}
    \right], 
\end{equation}
\begin{equation}
    \sigma^{\mathbf{e}}_i = \sqrt{\frac{1-\|R_i\|_2^2}{\|R_i\|_2(2-\|R_i\|_2^2)}}.
\end{equation}
For small scatter this reduces to the normal standard deviation.
The total error $\sigma_i$ on angle prediction $\theta_i$ for track $\mathbf{x_i}$ is then given by the quadrature combination
\begin{equation}
\label{eqn:toterr}
    \sigma_i^2 = \left(\frac{\sigma^{\mathbf{e}}_i}{2}\right)^2 + \frac{1}{M}\sum\limits_{j=1}^M\left(\frac{\sigma^{\mathbf{s}}_{ij}}{2}\right)^2, 
\end{equation}
where the factors of $1/2$ transform from errors on $2\theta_i$ to errors on $\theta_i$.

\begin{figure*}
\centering
\includegraphics[width=1.036\textwidth]{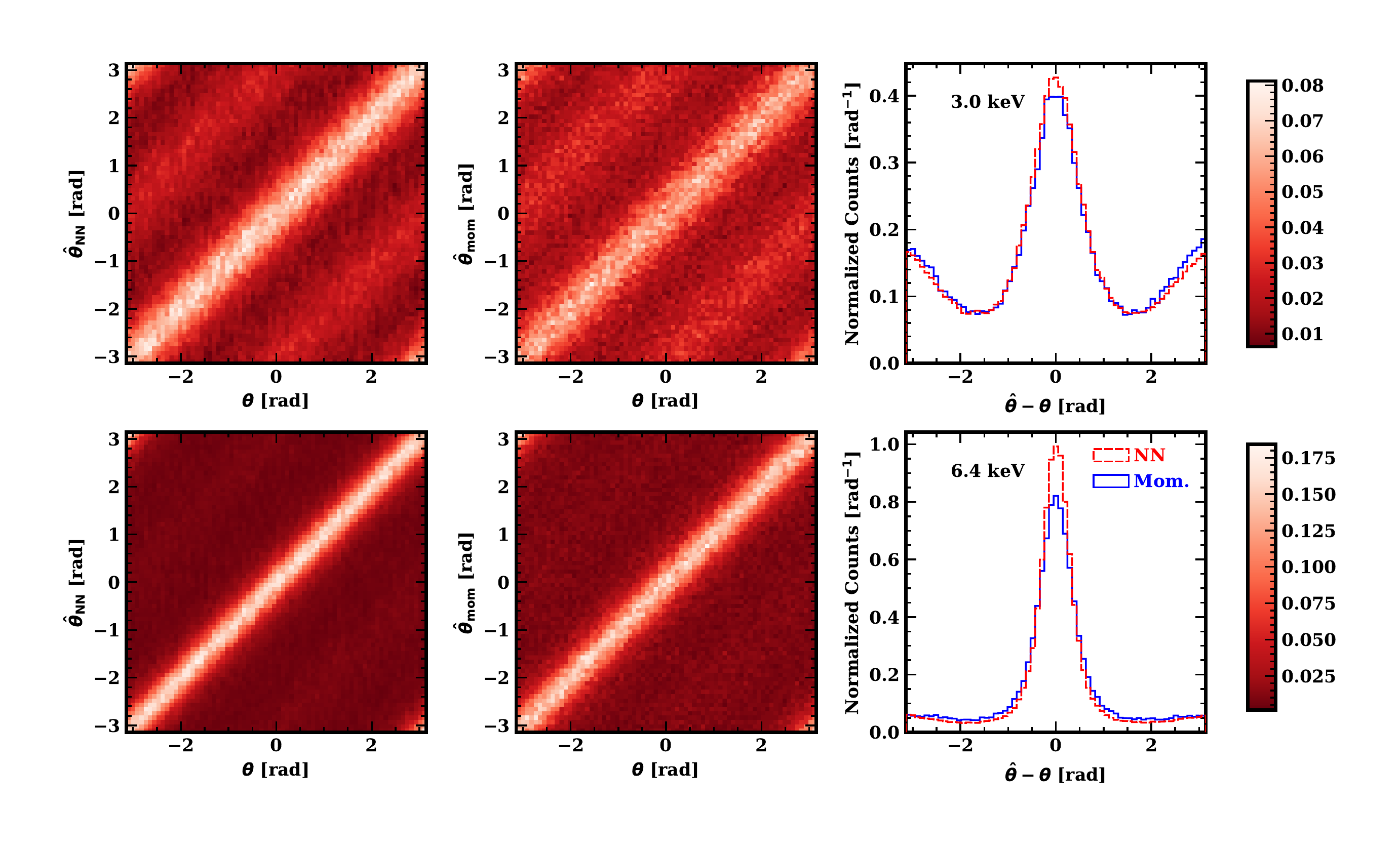}
\captionsetup{justification=centering,skip=2pt,font=scriptsize}
\caption{NN (left) and moments (middle) predicted photoelectron angle ${\hat \theta}$ {\it vs.} true photoelectron angle $\theta$ for 3.0 (top) and 6.4 (bottom) keV. Right: histograms of the angle differences. Our NN predictions have more measurements at the true angle (fewer $\pi$ offsets at low energy), and better recovery of the true $\theta$, especially at higher energies. Note that a $\pi$ ambiguity in prediction does not affect polarization measurement, which depends on $2\hat{\theta}$.}
\label{fig:confusion}
\end{figure*}

\begin{figure*}
\centering
\includegraphics[width=1.05\textwidth]{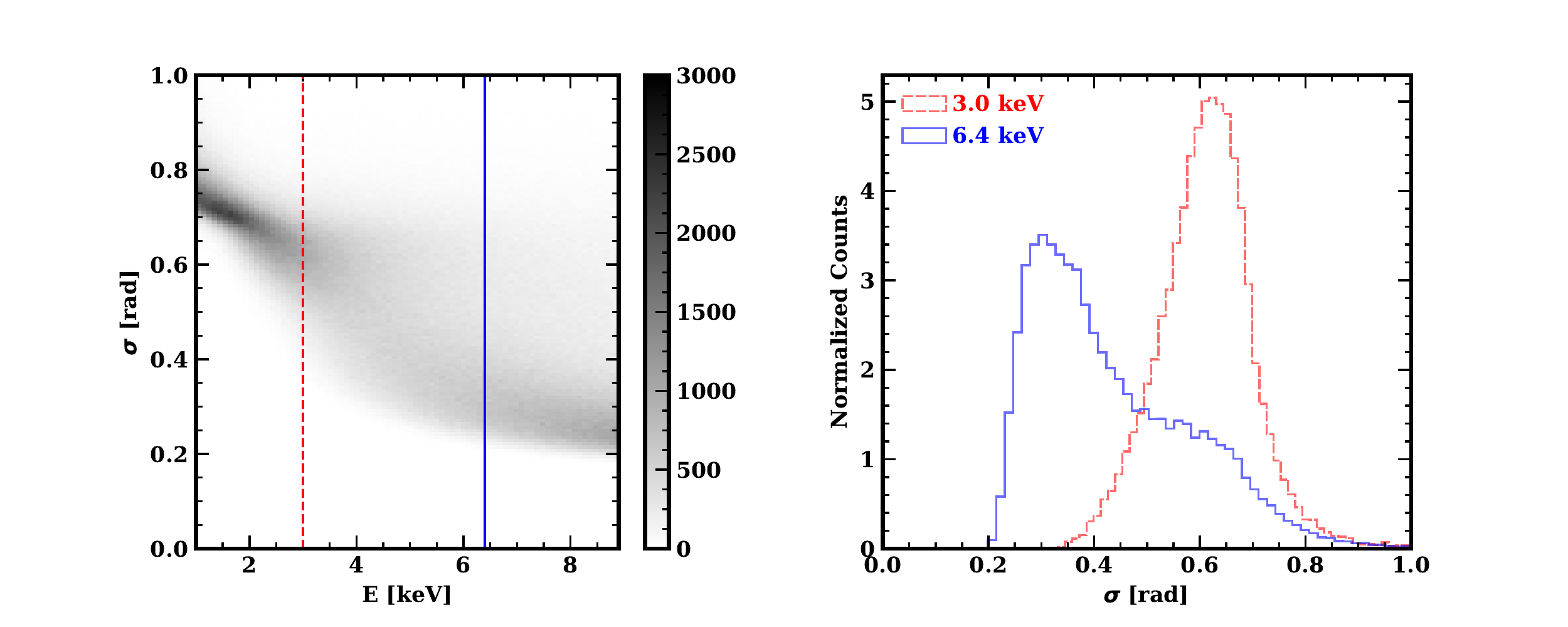}
\captionsetup{justification=centering,skip=2pt,font=scriptsize}
\caption{Distribution of deep ensemble predicted total errors $\sigma$ (eq.~\ref{eqn:toterr}) across the IXPE energy spectrum (left). The right-hand plot shows the $\sigma$ distribution for two specific energies $3.0$ keV (red), $6.4$ keV (blue). Low energy tracks tend to cluster around higher predictive uncertainty. High energy tracks have a tail of poorly predicted examples. Many of these correspond to the events converting outside the gas volume (see falling black curve in fig.~\ref{fig:mod}).}
\label{fig:errors}
\end{figure*}

\section{NN training and selection}

Here we describe the training procedure for individual NNs (\textsection 4.1) in the ensemble and the process of NN selection for ensemble membership.
\subsection{Data Sets}
Our training and validation data sets consist of simulated tracks (e.g.\,fig.~\ref{fig:hex}) generated via a Monte Carlo Geant4 \cite{agostinelli_geant4simulation_2003} simulation, part of IXPE's GPD software suite \citep{bellazzini_novel_2003}. The track energies uniformly span $1.0 - 9.0$ keV, IXPE's most sensitive range and are unpolarized (uniform track angle distribution). We simulate for the expected IXPE gas pressure of 687\,mbar. Each track is labelled with its 2D track angle vector $({\rm cos}\,\theta,{\rm sin\,}\theta)$, absorption point coordinates (on the square grid) $(x,y)$ and its energy $E$. This gives a final feature vector $({\rm cos}\,\theta,{\rm sin}\,\theta,x,y,E)$. 
We simulate 3.5 million tracks with a flat energy distribution. 
We split these 3.5 million tracks into a training set, validation set and test set where the validation set and test set make up $5\%$ of the total. Finally we have tested the performance on tracks from real GPDs; while we do not describe the detailed response here, we find polarization performance very similar to that realized from the simulated tracks.

\subsection{Training}

We use a ResNet-19 \citep{he_deep_2015} CNN architecture as our base NN. ResNets and their variants (like DenseNet \citep{huang_densely_2018}) are the current state of the art in image classification. They contain 'skip'-connections in between their layers that lead to faster and more robust training. This particular architecture is large enough to over-fit the training set, and trains in a reasonable amount of time ($\sim 15$ hours for 150 epochs on 4 Nvidia Titan GPUs, using a batch size of 2048). Before training we normalize the track images, subtracting the pixel-wise mean from each track image and dividing by the pixel-wise standard deviation (where the mean and standard deviation are calculated over the full training set). The track energy and absorption point labels are similarly processed.
Normalizing the training data helps prevent vanishing and exploding gradients during the NN training procedure and lead to faster convergence. We use stochastic gradient descent with momentum as our optimizing algorithm, typical in computer vision tasks \citep{sutskever_importance_2013}, with a stepped decaying learning rate starting at 0.01.
We choose batch sizes of 512, 1024, 2048 tracks. The training procedure seeks to minimize the loss function given by eq. \ref{eqn:DE_tot_loss} over the NN parameters. We tune the hyperparameters $\alpha,\beta,\gamma,\delta,\epsilon$ to minimize the MSE on the predicted track angle for the validation set, while retaining energy and absorption point accuracy. We find, typically, $\alpha \sim 1$, $\beta \sim 0.2$, $\gamma\sim 0.2$, $\delta \sim 5 \times 10^{-5}$ and $\epsilon \sim 0.5$ keV  work well for angle accuracy and energy resolution. Absorption point and energy resolution depend weakly on the choice of $\beta$ and $\gamma$.
The values of $\beta$ and $\gamma$ are $<1$, reflecting our choice to focus on the measurement of track angles. 

\subsection{Selection}
For our deep ensemble, we train 27 individual NNs in the manner described above -- 9 NNs for each of the 3 different batch sizes (encouraging model diversity). We use the entire training data set for each NN since deep NNs typically perform better with more data. After training these individual NNs, we down-select to the best performing 10, as measured by the track angle MSE loss $L_{\theta}$ (eq.~\ref{eqn:DE_phi_loss}) on the reserved test set. For the test set all three rectified track image pairs are evaluated, and the resulting track angle predictions are rotated back, to remove any imprinted hexagonal angle prediction bias. Since we save a checkpoint every 10 epochs during the training, we evaluate the NNs at all middle-late stages of their training in our selection, retaining the best stage. This form of early-stopping helps prevent over-fitting of the training set. Fig.~\ref{fig:confusion} shows the photoelectron angle recovery for an ensemble of NNs and the moment analysis at two example energies, 3.0 and 6.4 keV.

In principle, absorption point and energy resolution errors could be factored into the NN down-selection, but since polarization is our principal interest (and the most difficult to train), we focus on $L_{\theta}$. The results in \textsection5 obtain from ensembles of the 10 networks with the lowest $L_{\theta}$.

\section{Polarization estimation}

Unlike many typical deep learning problems, polarization estimation requires predicting a distribution measurement from a large number of events. We have split the problem in two, first extracting features via NNs (\textsection2), then forming the required measurement. Here we combine our deep ensemble's feature predictions to produce best estimates for the polarization fraction $\Pi$, EVPA $\phi$ and their errors.

The basic problem is to estimate $\Pi$ and $\phi$ from a set of measured track angles $\{\theta_i\}^N_{i=1}$. As described in the introduction, the track angles exhibit a sinusoidal modulation with period $\pi$
\begin{equation} \label{eq:likelihood}
    p(\theta|\mu,\phi) = \frac{1}{2\pi}\big(1 + \mu{\rm cos}\big[2(\theta - \phi)\big]\big)
\end{equation}
where $0 \leq \mu \leq 1$, $-\pi/2 \leq \phi < \pi/2$ and $-\pi \leq \theta < \pi$. The intrinsic polarization fraction is then given by $\Pi = \mu / \mu_{100}$. While simple analyses often bin the data set $\{\theta_i\}^N_{i=1}$ and then fit to estimate $\mu,\phi$, this provides an inevitable loss of information, and poor performance when the bin counts are limited. Maximum likelihood methods are preferable, being unbinned. \citet{kislat_analyzing_2015} have developed an unbinned method, working with total Stokes parameters. They define
\begin{subequations}
\label{eqn:stokes}
\begin{eqnarray}
 Q = 2\sum\limits_{i=1}^N {\rm cos}2\theta_i \\
 U = 2\sum\limits_{i=1}^N {\rm sin}2\theta_i 
\end{eqnarray}
\end{subequations}
where $N$ is total number of track angles $\theta_i$. Then $\mu,\phi$ can be simply calculated as
\begin{subequations}
\label{eqn:stokes_final}
\begin{eqnarray}
    &\mu =& \frac{1}{N}\sqrt{\big(Q^2 + U^2\big)} \\
    &\phi =& \frac{1}{2}{\rm arctan}\frac{U}{Q}
\end{eqnarray}
\end{subequations}
The strength of this kind of analysis is that the errors on the Stokes parameter are easily computed and well behaved. This method is unbiased and faster than a full maximum likelihood fit. 

However, the different events can give quite varied constraining power. For example, low-energy short tracks inevitably have poorly constrained $\theta$ (and, given the typical astrophysical energy spectrum and the detector response, most events will have low energy!). A subset of events with higher energy or cleaner, longer tracks may be most useful. For this reason it is essential to incorporate some form of quality control in the tracks we use for our polarization estimates. Traditionally one applies track cuts, but this is a sub-optimal since one can throw away a large fraction of the data. A weighting scheme should be preferred.

It is possible to incorporate event weights into the Stokes parameters method by defining:
\begin{subequations}
\label{eqn:weightstokes}
\begin{eqnarray}
 &Q =& 2\sum\limits_{i=1}^N w_i{\rm cos}2\theta_i, \\
 &U =& 2\sum\limits_{i=1}^N w_i{\rm sin}2\theta_i, \\
 &I =& \sum\limits_{i=1}^N w_i
\end{eqnarray}
\end{subequations}
where $w_i$ is the relative weight for the event angle $\theta_i$ and $I$ is used instead of $N$ in eq.~\ref{eqn:stokes_final}. 

Fig.~\ref{fig:errors} shows the distribution of NN predicted total errors $\sigma$ for track data sets spanning all energies. These are equivalent to the track reconstruction quality. Even high energy data sets show tails of poorly reconstructed tracks. 

\subsection{Importance weighted maximum likelihood}

Our deep NN ensembles provide us with statistical and epistemic error estimates for each event. Our approach is to incorporate these into our measurement of the polarization parameters. In maximum likelihood estimation, the negative log-likelihood of the polarization parameters given the measured track angles is minimized over the likelihood domain:
\begin{equation}
    \begin{array}{ll}
    \underset{{\rm over}\mu,\phi}{\mbox{minimize}}   &  -\sum\limits_{i=1}^N {\rm log}\big(1 + \mu{\rm cos}\big(2(\theta_i - \phi)\big)\big) \\
    \mbox{subject to} & 0 \leq \mu \leq 1 \\
    & -\pi/2 \leq \phi < \pi/2,
    \end{array}
    \label{eqn:min}
\end{equation}
where eq.~\ref{eq:likelihood} is used as the likelihood function $L(\{\theta_i\}|\mu,\phi)$ and we have dropped constant terms, since they do not affect the minimization. The optimal values of eq.~\ref{eqn:min}, $(\mu^{\star},\phi^{\star})$, are the final polarization estimates. An approximation of errors on these estimates (or equivalently on the Stokes fluxes from the source) can be calculated analytically using Fisher information. We make the substitution $\mu = \Pi\mu_{100}$ to get 
\begin{subequations}
\begin{eqnarray}
\label{eqn:err_pi}
\sigma(\Pi) \approx \frac{1}{\mu_{100}}\sqrt{\frac{2-(\Pi\mu_{100})^2}{(N-1)}} \\
\label{eqn:err_phi}
\sigma(\phi) \approx \frac{1}{\Pi\mu_{100}\sqrt{2(N-1)}}. 
\end{eqnarray}
\end{subequations}
Note both errors depend strongly on the modulation factor $\mu_{100}$, with higher modulation factors leading to smaller prediction errors. There is also a dependence on the total number of measured tracks $N$, making extensive track cuts detrimental.

We modify equation~\ref{eqn:min} to an importance-weighted maximum likelihood estimate
\begin{equation}
    \begin{array}{ll}
    \underset{{\rm over} \mu,\phi}{\mbox{minimize}}   &  -\sum\limits_{j=1}^M\sum\limits_{i=1}^N \sigma^{-\lambda}_{ij}{\rm log}\big(1 + \mu{\rm cos}\big(2(\theta_{ij} - \phi)\big)\big) \\
    \mbox{subject to} & 0 \leq \mu \leq 1 \\
    & -\pi/2 \leq \phi < \pi/2,
    \end{array}
    \label{eqn:e-opt-prob}
\end{equation}
where $M$ is the number of NNs in the deep ensemble, $\theta_{ij}$ are the predicted track angles and $\sigma_{ij} = \sqrt{(\sigma^{\mathbf{s}}_{ij}/2)^2 + (\sigma^{\mathbf{e}}_{i}/2)^2}$ are the total predicted errors on $\theta_{ij}$ (\textsection2.3). The $j$th NN in the ensemble makes a prediction ($\theta_{ij},\sigma_{ij}$) for the $i$th track. 
The predicted errors $\sigma_{ij}^{-1}$ are used as importance weights \citep{karampatziakis_online_2011, hu_weighted_2002}, so that low $\sigma$ tracks are weighted more strongly. The parameter $\lambda$ controls a simple weighting scheme: for $\lambda = 0$ the log-likelihood is unweighted and we recover the Stokes method (Equations 9 and 10). With high $\lambda$ the best-measured tracks dominate the polarization estimate. This increases $\mu_{100}$ but leads to larger fluctuations in the estimated values, as well as increased sensitivity to reconstruction biases (e.g.\,when $\theta$ aligns with the underlying hexagonal grid, giving cleaner tracks). In \textsection4.2 we provide a prescription for selecting $\lambda$. The best $\lambda$ will depend on the source spectrum.

We can rewrite eq.~\ref{eqn:e-opt-prob} as a convex optimization problem:
\begin{equation}
\label{eqn:convex}
    \begin{array}{ll}
    \underset{{\rm over}\, \mathbf{x}}{\mbox{minimize}}   &  -\sum\limits_{j=1}^M\sum\limits_{i=1}^N \sigma^{-\lambda}_{ij}{\rm log}\big(1 + \mathbf{v}_{ij}^{T}\mathbf{x}\big) \\
    \mbox{subject to} & \|\mathbf{x}\|_2 \leq 1
    \end{array}
\end{equation}
where $\mathbf{v}_{ij} = ({\rm cos}\theta_{ij},{\rm sin}\theta_{ij})$ and $\mathbf{x} = (\mu{\rm cos}\phi,\mu{\rm sin}\phi)$. By recasting eq.~\ref{eqn:e-opt-prob} as a convex optimization problem, we have a guaranteed globally optimal solution for $(\mu,\phi)$. We can solve eq.~\ref{eqn:convex} quickly and efficiently using second order Newton methods. In practice we use the robust open source software Ipopt \cite{wachter_implementation_2006}.

When we include the full distribution of $M$ track angle predictions, we most accurately approximate the true model uncertainty. However, it is possible to simplify the analysis by averaging the track angle predictions $\theta_{ij}$ over $j$:
\begin{equation}
\theta_i = \frac{1}{2}{\rm atan2}\Big(\frac{1}{M}{\sum\limits_{j=1}^{M}{\rm sin}2\theta_{ij}},\frac{1}{M}{\sum\limits_{j=1}^{M}{\rm cos}2\theta_{ij}}\Big)
\end{equation}
and using $\sigma_i$ (eq.~\ref{eqn:toterr}) for the total error on $\theta_i$. This allows the importance weighted maximum likelihood to recover the weighted Stokes analysis (eqs.~\ref{eqn:stokes_final}-\ref{eqn:weightstokes}), where $w_i = \sigma^{-\lambda}_i$. This approximation is much easier to compute, but does not fully exploit the epistemic uncertainty.


\subsection{Figure of merit}

We require a figure-of-merit (FoM) for the polarization measurement, a scalar value representing the signal to noise ratio of a given detector and analysis scheme. The standard figure-of-merit used in X-ray polarimetry currently is the minimum detectable polarization (MDP) \citep{weisskopf_understanding_2010}. MDP$_{99}$ is the polarization fraction that has a 1\% probability of being exceeded by chance for an unpolarized source. The probability $p(\mu,\phi)$ of measuring modulation $\mu$ and EVPA $\phi$ given that the true modulation and phase are $\mu_0$ and $\phi_0$ is \citep{weisskopf_understanding_2010}
\begin{equation}
\label{eqn:err_prob}
p(\mu,\phi) = \frac{N\mu}{4\pi}{\rm exp}\bigg(-\frac{N}{4}\big[\mu^2 + \mu_0^2 - 2\mu\mu_0{\rm cos}2(\phi - \phi_0)\big]\bigg).    
\end{equation}
Here $N$ is the total number of measured tracks as usual. The MDP$_{99}$ may be found by integrating eq.~\ref{eqn:err_prob} for $\mu_0 = 0$, resulting in 
\begin{equation}
\label{eqn:MDP}
{\rm MDP}_{99} \approx \frac{4.29}{\mu_{100}\sqrt{N}}
\end{equation}
where $\mu_{100}$ accounts for imperfect polarization recovery; the source polarization will be $\Pi = \mu / \mu_{100}$. Note the similarity to eq.~\ref{eqn:err_pi}: MDP is effectively a (inverse) ratio of recovered signal $\mu_{100}$ to noise $\sim 1 / \sqrt{N}$. For a given detector, track reconstruction algorithms with lower MDPs are better. 

For a weighted polarization estimate, the error on recovered $\Pi,\phi$, and equivalently the noise denominator term $\sqrt{N}$, is no longer given by the total number of tracks since some tracks can contribute significantly more than others. We define $N_{\rm eff}$ to replace $N$ in the MDP$_{99}$, and in eqs.~\ref{eqn:err_phi}, for a weighted scheme.
For the weighted Stokes method (eqs.~\ref{eqn:weightstokes}),
\begin{equation}
\label{eqn:neff}
    N_{\rm eff} = I^2 / \sum\limits^N_{i=1}w_i^2.
\end{equation}
For an importance weighted maximum likelihood incorporating all $M$ NN ensemble predictions (eq.~\ref{eqn:e-opt-prob}) there is no closed form solution for $N_{\rm eff}$. However, with numerical simulations we can study the variance of our importance weighted maximum likelihood results and measure the effective $\mu_0$, if any, from the offset and $N_{\rm eff}$ from the width. This allows us to use MDP$_{99}$ as our 'Figure of Merit' for comparing analysis schemes, including cuts and weights, when we replace $N$ by $N_{\rm eff}$ in Eq.~\ref{eqn:MDP}, where we determine $N_{\rm eff}$ by fitting Eq.\,\ref{eqn:err_prob} to a bootstrap-sampled distribution of $(\mu,\phi)$ for an unpolarized source. In this fit, we maximize the likelihood $p(\{\mu,\phi\}\mid N, \mu_0, \phi_0)$, where $\{\mu,\phi\}$ is our set of bootstrap samples. The optimal value $(N^{\star}, \mu_0^{\star}, \phi_0^{\star})$ defines $N_{\rm eff} = N^{\star}$. We do not include the estimator bias $\mu_0^{\star}$ directly in the FoM because any useful estimator should either have a negligible bias or the bias, if known, should be divided out (as in \citet{kitaguchi_convolutional_2019}); $\mu_0^{\star}$ is simply a tool to evaluate any residual bias in the track reconstruction. Indeed, we find that the $\mu_0^{\star}$ are small, consistent with the residual statistical polarization in our finite-sized event sets.

\begin{figure}
\centering
\includegraphics[width=0.65\textwidth]{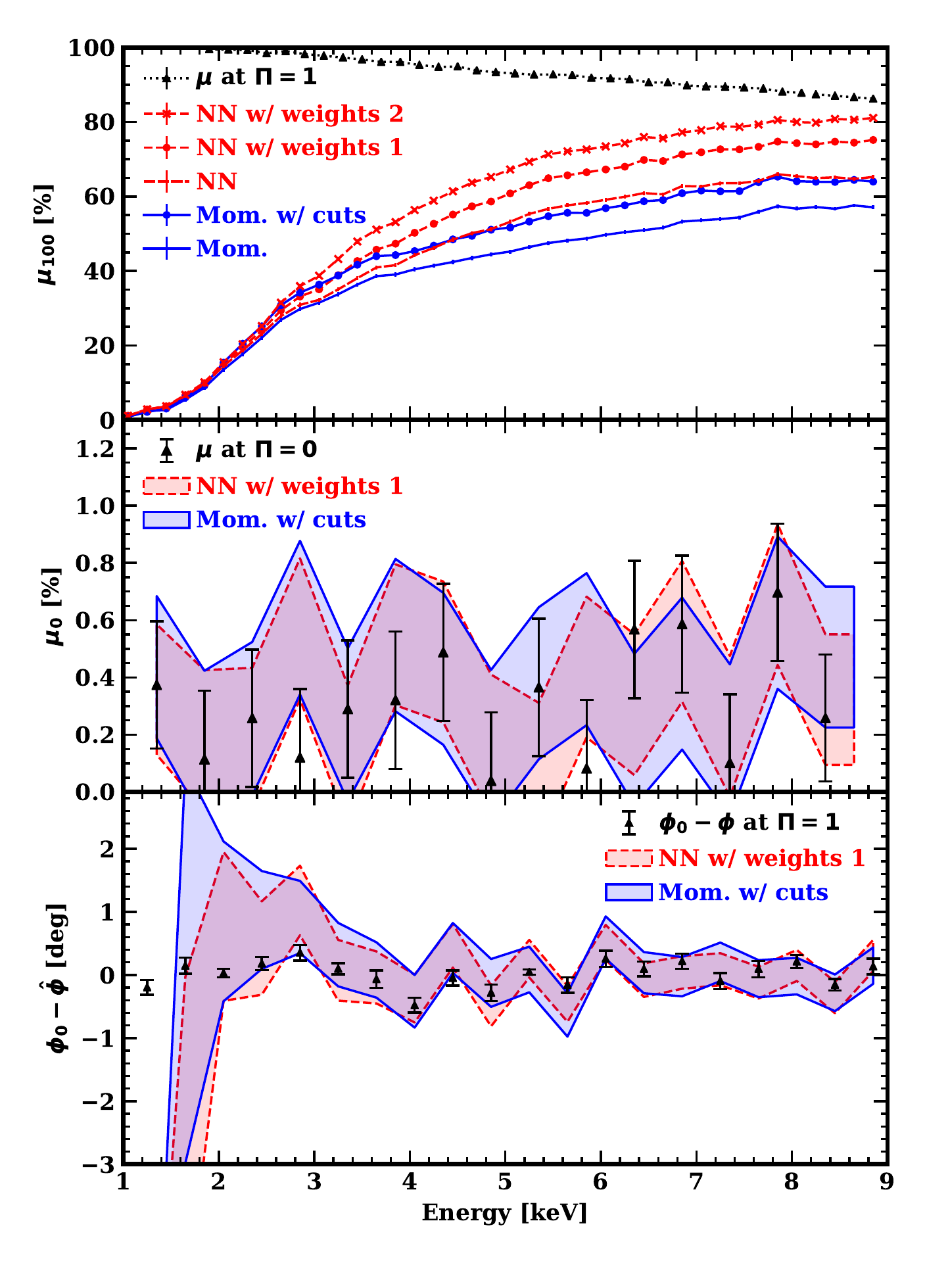}
\vskip -5mm
\caption{Modulation response for analysis of simulated GPD tracks. Top: The available signal $\mu$ for a 100\% polarized source, after dilution by window/GEM conversion events (black dotted line). Standard moments analysis and moments with a 20\% ellipticity cuts are compared with the deep ensemble NN analysis with uniform weighting. The NN $\mu$ improves using $\lambda = 1$ (dots) and $\lambda = 2$ (crosses) weights. 
Middle: Response to an unpolarized signal. Triangular points show the residual polarization of the $\sim3.4\times10^{5}$ event simulation, with $1\sigma$ error bars. Solid (moments with 20\% cuts) and dashed (NN $\lambda=1$) bands show the recovered signal $1\sigma$ statsitical uncertainty ranges.
Bottom: Recovered EVPA error for a polarized signal. Triangular error bars show the residual EVPA from the finite simulation.}
\label{fig:mod}
\end{figure}

Bootstrapping works for any algorithm (when one can compute sufficient bootstrap samples), so this FoM can be used for all polarization estimation methods discussed, with or without event cuts or weights. 
For a given set of tracks, the weighting scheme $\lambda$ for our importance weighted MLE (eq.~\ref{eqn:e-opt-prob}) is chosen by evaluating the MDP$_{99}$ for a set of $\lambda$s and choosing the best performer. 
Using the weighted Stokes approximation -- where prediction from the $M$ NNs in the ensemble are averaged for each track -- the best $\lambda$ is faster to compute since $N_{\rm eff}$ is given analytically (eq.~\ref{eqn:neff}); no bootstrapping is required.
In \textsection5 we identify the best $\lambda$ using the weighted Stokes approximation first, then apply the full importance weighted MLE, using the bootstrap analysis to find the final MDP$_{99}$.
We show examples of the bootstrap fits in \textsection5.

\begin{figure}
\centering
 \includegraphics[width=1.0\textwidth]{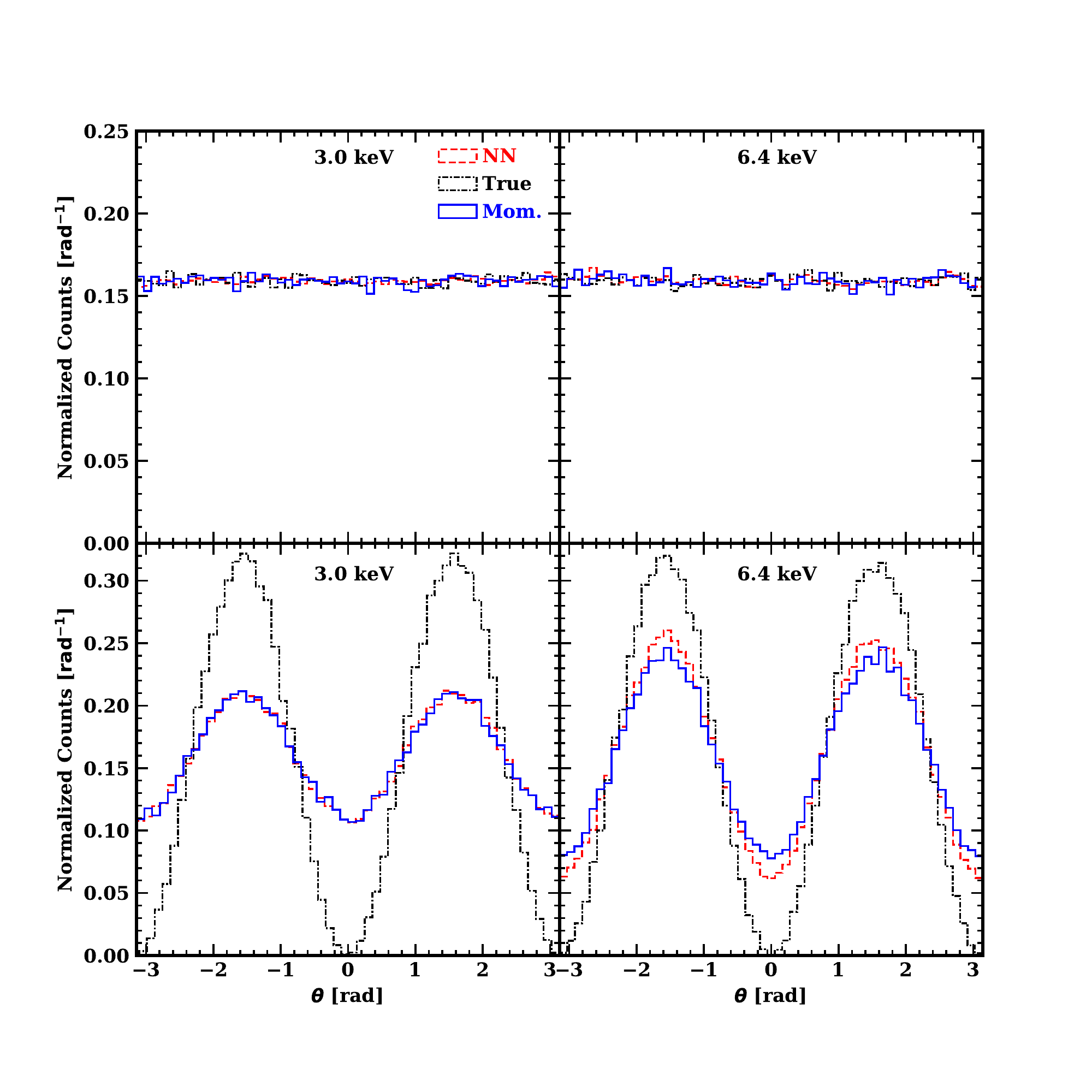}
\vskip -10mm
\caption{Track angle recovery for unpolarized (top row) and polarized (bottom row) simulated data for $3.0$ and $6.4$\,keV. The original photoelectron angle distribution is shown in black; standard moment analysis reconstruction is in blue and unweighted NNs in red. The top panels show negligible residual polarization, the bottom panels show lack of bias and a modest increase in NN sensitivity, especially at higher energies. }
\label{fig:4hist}
\end{figure}

\section{Results}

\begin{figure}
\centering
\includegraphics[width=0.6\textwidth]{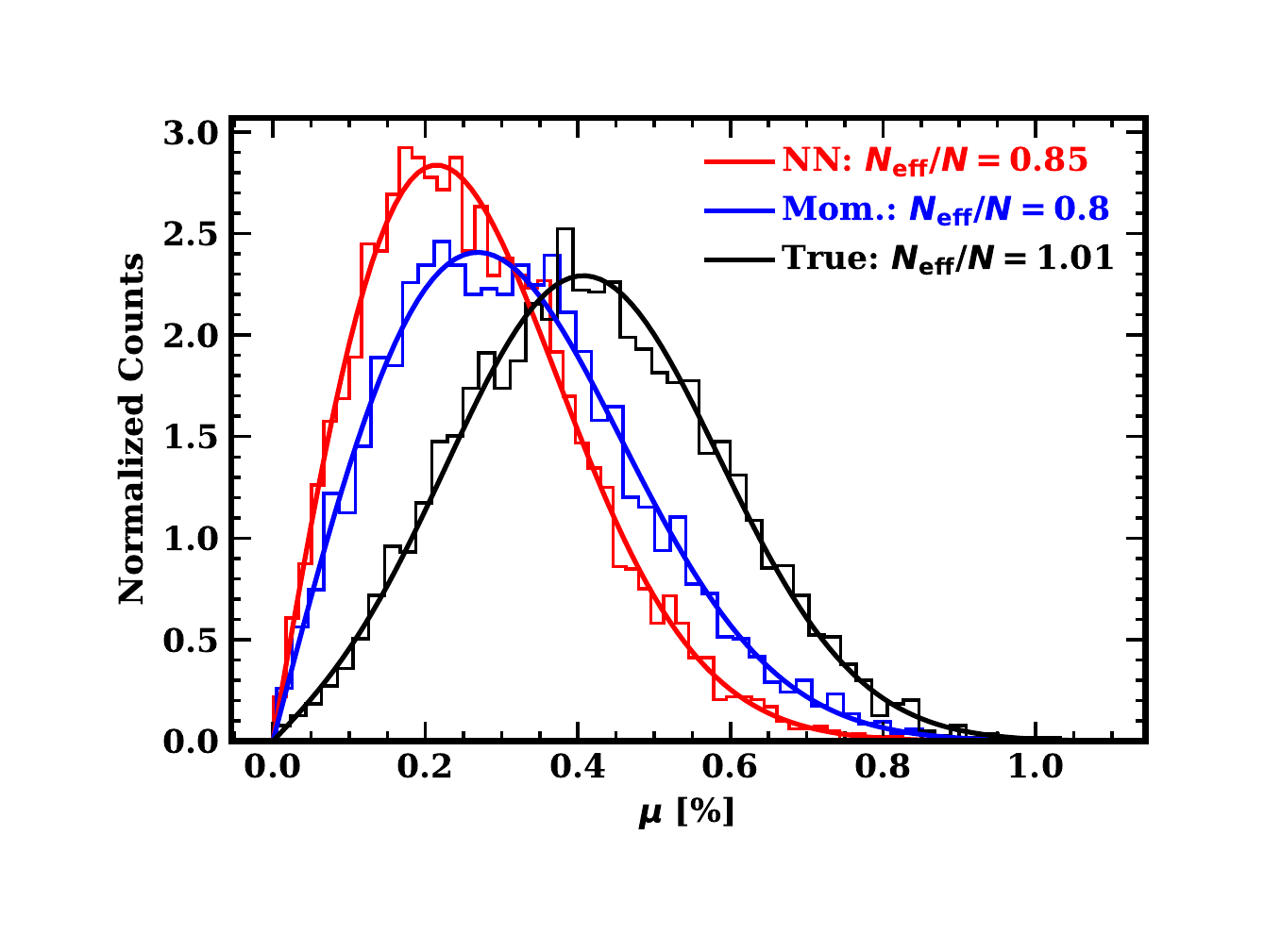}

\caption{Bootstrapped modulation distributions for a PL2 test data set (0.6 million simulated tracks with 5000 bootstrap samples). The residual modulation from the original simulated photoelectron directions are the black histogram, weighted NNs $\lambda = 1.83$ are in red and moment analysis with standard cuts in blue. The solid lines show the fits of Eq.\,\ref{eqn:err_prob} to these distributions (the fitting also includes the bootstrapping distribution on $\phi$, not shown here). From these fits we get an estimate of $N_{\rm eff}$ for each method to use in the FoM (Eq.\,\ref{eqn:MDP}). For weighted NN, the result gives a slightly improved MDP$_{99}$, as listed in Table~\ref{tab:fom}.}
\label{fig:boot}
\end{figure}

All of the training, testing and validation have been performed on the simulated event track images. These have been used to train NNs and down-select to an ensemble having maximum sensitivity and minimum residual error.  We have performed limited tests with real GPD data and confirm that we can obtain comparable accuracy with these images, but unsimulated physical effects in the GPD hardware and the peculiarities of individual flight GPD will require networks tuned for specific devices. Here we focus on the general performance.

\subsection{Polarization}

To maximize polarization sensitivity, the appropriate metric for performance is the FoM defined in \textsection4.2. However, it is instructive to examine the $\mu_{100}$ and unpolarized $\mu_0$ energy dependence across the IXPE spectral range. We show these results for simulated data in fig.~\ref{fig:mod} for our method with and without weights, and the moment analysis with and without cuts. In the upper panel, the black line at top shows the modulated signal from a simulated data set corresponding to a 100\% polarized ($\Pi=1$) source. The decrease toward high energy represents the increasing fraction of events which convert in the detector window or the GEM foil. These are truncated and/or highly scattered, with a fraction of the normal gas energy deposition and retain little polarization information. Thus perfect reconstruction can at best reach the black line. The blue solid line shows the performance of the mission default (Moments) analysis. If one knows the event energy {\it a priori} (e.g.\,for a calibration source) one can cut on recovered energy to remove many of the incomplete tracks, improving the polarization response (blue dotted line). For astrophysical sources with unknown event energies, this simple effective prescription cannot be applied. More sophisticated shape cuts could in principle recover a fraction of this improvement -- however one must recall that these cuts decrease the sample size and when considering the polarization sensitivity of a given data set, this substantially reduces the gains from the cuts.

\begin{table}[t!]
\centering
\begin{tabular}{@{}l l l l l@{}}
\toprule
{\textbf{Method}}&$\mu_{100}$(\%)&$N_{\rm eff}/N$&$\lambda$&MDP$_{99}$(\%) \\
\midrule
\textcolor{blue}{\bf Mom.}& 27.0 & 1.0 & 0 &\textcolor{blue}{5.03 $\pm$ 0.02}\\ 
\textcolor{blue}{\bf Mom. w/ cut}& 31.3 & 0.796 & -- &\textcolor{blue}{4.88 $\pm$ 0.03}\\ 
\textcolor{blue}{\bf Mom. w/ weights}& 31.4 & 0.878 & 0.67 &\textcolor{blue}{4.61 $\pm$ 0.01 $\leftarrow$}\\ 
\midrule
\textcolor{red}{\bf NN }& 28.7 & 1.0 & 0 &\textcolor{red}{4.72 $\pm$ 0.02}\\
\textcolor{red}{\bf NN w/ weights}& 32.6 & 0.954 & 1  &\textcolor{red}{4.26 $\pm$ 0.02}\\ 
\textcolor{red}{\bf NN w/ weights}& 36.8 & 0.812 & 1.83 &\textcolor{red}{4.09 $\pm$ 0.02 $\leftarrow$}\\ 
\textcolor{red}{\bf NN w/ weights (bootstrap)}& 36.1 & 0.852 & 1.83 &\textcolor{red}{4.07 $\pm$ 0.02 $\leftarrow$}\\ 
\textcolor{red}{\bf NN w/ weights}& 37.7 & 0.763 & 2  &\textcolor{red}{4.12 $\pm$ 0.02}\\ 
 \bottomrule
\end{tabular}
\caption{Sensitivity analysis for $10^5$ 2-8\,keV photons with a $dN/dE \sim E^{-2}$ spectrum and {\it IXPE}'s energy response. ${\rm MDP}_{99}$ gives the sensitivities for the various cuts and weights; smaller MDP$_{99}$ is better. The arrows show the minimum MDP$_{99}$ found by optimizing over $\lambda$. With a bootstrap analysis (fig.~\ref{fig:boot}), the weighting takes better advantage of the epistemic uncertainty for a small additional MDP$_{99}$ decrease at this $\lambda$.}
\label{tab:fom}
\end{table}

\begin{table}[t!]
\centering
\begin{tabular}{@{}l l l l l@{}}
\toprule
{\textbf{Method}}&$\mu_{100}$(\%)&$N_{\rm eff}/N$&$\lambda$&MDP$_{99}$(\%) \\
\midrule
\textcolor{blue}{\bf Mom.}& 29.9 & 1.0 & 0 &\textcolor{blue}{4.55 $\pm$ 0.02}\\ 
\textcolor{blue}{\bf Mom. w/ cut}& 34.7 & 0.804 & --  &\textcolor{blue}{4.36 $\pm$ 0.01}\\ 
\textcolor{blue}{\bf Mom. w/ weights}& 35.1 & 0.867 & 0.60  &\textcolor{blue}{4.15 $\pm$ 0.01 $\leftarrow$}\\ 
\midrule
\textcolor{red}{\bf NN}& 32.1 & 1.0 & 0 &\textcolor{red}{4.22 $\pm$ 0.02}\\ 
\textcolor{red}{\bf NN w/ weights}& 37.3 & 0.935 & 1  &\textcolor{red}{3.76 $\pm$ 0.01}\\ 
\textcolor{red}{\bf NN w/ weights}& 41.2 & 0.812 & 1.65  &\textcolor{red}{3.65 $\pm$ 0.01 $\leftarrow$}\\ 
\textcolor{red}{\bf NN w/ weights (bootstrap)}& 40.9 & 0.835 & 1.65  &\textcolor{red}{3.63 $\pm$ 0.01 $\leftarrow$}\\ 
\textcolor{red}{\bf NN w/ weights}& 43.6 & 0.707 & 2  &\textcolor{red}{3.70 $\pm$ 0.01}\\ 
 \bottomrule
\end{tabular}
\caption{Sensitivity analysis, as for Table \ref{tab:fom}, but for $10^5$ 2-8\,keV photons with a $dN/dE \sim E^{-1}$ spectrum and {\it IXPE}'s energy response.} 
\label{tab:fom1}
\end{table}

The performance of our initial NN analysis (red dashed line) is better than the moments analysis and at high energies, even matches moments after the ellipticity cuts. Of course, we can use our event quality metric to make a better weighted measurements (red starred and dotted lines) and this substantially increases the modulation sensitivity, at a modest cost in the effective number of sample events.

While high polarization sensitivity $\mu_{100}$ is desired, it is essential that the analysis chain not induce spurious signals from unpolarized sources. The middle panel summarizes tests for this effect. The finite size of the simulated unpolarized data set guarantees a residual statistical polarization. This is shown by the black triangles; an ideal measurement should not induce polarization significantly in excess of this value. The colored traces show the $1\sigma$ range for weighted NNs and the moment analysis with cuts. We see that the NN analysis induces no significant polarization and, like the moments analysis, meets mission requirements at all energies. The slightly larger $1\sigma$ range on the moment analysis with cuts are a marker for increased measurement noise -- the trade off between increased signal vs. noise is best evaluated with our FoM.
Here we are measuring a small residual signal so the energy bins are larger, containing $\sim 340,000$ test tracks.

We can compare with the PRAXyS NN simulation results of \citet{kitaguchi_convolutional_2019} for the non-imaging low pressure PRAXyS detector. While their $\mu_{100}$ sensitivity (at three energies) is roughly at the position of our $\lambda=1$ (red dotted-dashed) curve, their unpolarized signal had large systematics. In part this is because their networks had very large prediction biases (large $\mu_0$ in the middle panel) -- they were forced to divide out the $\theta$ pattern of these biases to obtain acceptable $\sim$1\% polarization levels. Our NN ensemble produces negligible bias at these energies, so we did not have to perform such a normalization, yet our residual polarization is substantially lower at all energies for the same number of test tracks.

\begin{figure}%
    \centering
    \subfloat{{\includegraphics[width=7.7cm]{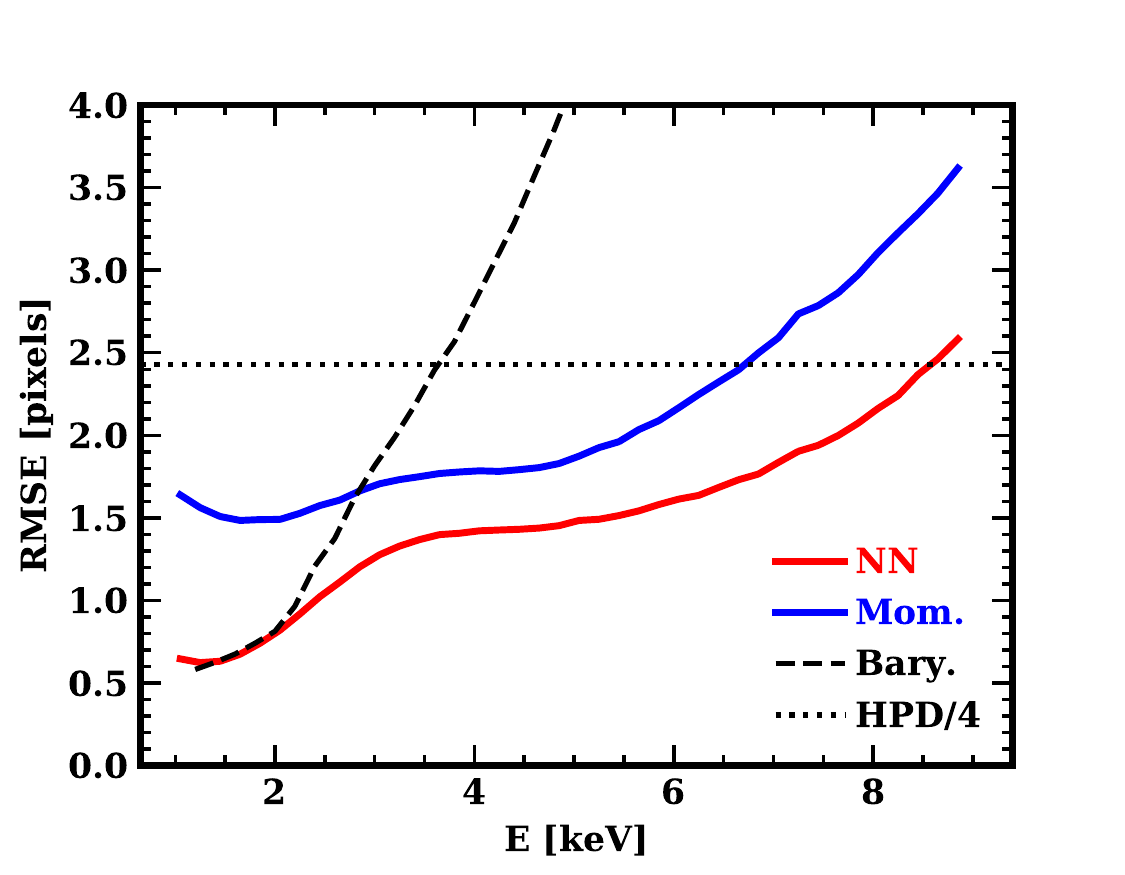} }}%
    \qquad
    \subfloat{{\includegraphics[width=7.9cm]{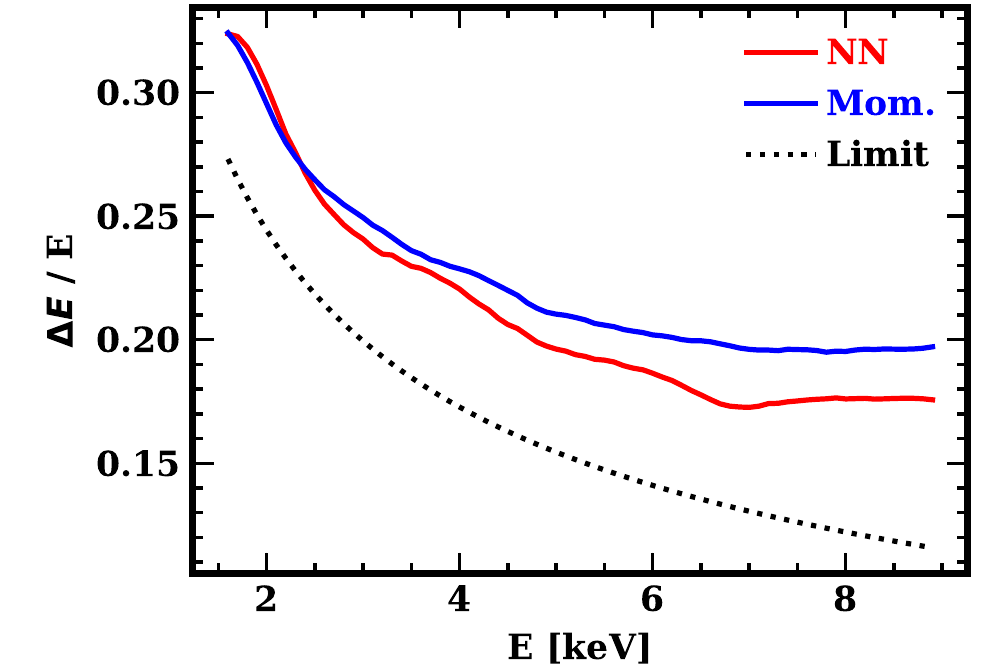} }}%
    \caption{Left: Photon absorption point localization. The NN analysis (red) does appreciably better than the moments analysis and matches the barycenter estimate at the lowest energies. All methods are adequate, as localization is much better than the size of the point source image produced by IXPE's mirrors. Right: Fractional width of the event energy estimate. The distributions are measured with the Median Absolute Deviation (MAD) statistic, which is less sensitive to the long tail from partial tracks (see Fig.\,\ref{fig:Eres1}). Values are converted as FWHM=3.46MAD, appropriate for a Gaussian peak. The NN analysis does slightly better than the standard PI count summing, especially at high energy. The theoretical resolution limit for the GPD Fano factor is shown by the dotted line.}%
    \label{fig:example}%
\end{figure}


The bottom panel of Fig.~\ref{fig:mod} shows the quality of EVPA recovery for the $100\%$ polarized case. Again NN performance is comparable to moments in $\phi$ recovery. Both methods have somewhat increased error at low energy, due to low recovered $\mu_{100}$ and limited counts analyzed in these energy bins
(eq.~\ref{eqn:err_phi}). For $\mu \sim 0$, $\phi$ becomes undefined.


To help visualize the reconstruction accuracy, we show the binned modulation curves for 0\% and 100\% polarization for two energies in fig.~\ref{fig:4hist}. Recall that we do {\it not} simply fit these histograms to recover polarization parameters (as in some less sophisticated analyses). Nevertheless, these are useful to show the lack of bias in our angle predictions. Some ML polarization analyses suffer from strong prediction bias, producing imperfect $2\pi$ symmetry and narrow peaks in the angular distribution, e.g. \citet{moriakov_inferring_2020}; no such artifacts are evident in our reconstructions. We can see that the unpolarized distributions for our NN method are as flat as those for the moments, but the NNs recover significantly more modulation in the polarized data. 

Fig.~\ref{fig:boot} shows the bootstrap distributions used to calculate $N_{\rm eff}$ for the full importance weighted maximum likelihood method.

\sloppy Tables \ref{tab:fom} and \ref{tab:fom1} show our FoM results for two power law datasets and several values of the importance weighting control parameter $\lambda$. These source spectrum power laws PL1 ($dN/dE = A E^{-1}$), and PL2 ($dN/dE = A E^{-2}$) extend across the IXPE energy spectrum. They are convolved with IXPE's effective area function \citep{weisskopf_imaging_2016}, to provide event energy distributions expected for real astrophysical sources. We include a moment analysis weighted by track ellipticities for comparison, although this analysis is not currently used by IXPE. We compute modulations for the 2-8\,keV events only, with values normalized to $10^5$ events in this range, as achievable for a moderately bright X-ray source.

In Table \ref{tab:fom} optimal $\lambda$ weighted NNs outperform the uncut moment analysis by $\sim 1.25\times$ in this FoM, and by $\sim 1.20\times$ with ellipticity cuts or $\sim 1.14\times$ with ellipticity weights for both PL2 and PL1 (tables \ref{tab:fom} and \ref{tab:fom1}). NN improvements in PL1 are slightly higher than PL2 since there are more high energy tracks. 
Note also that the improvement of our method comes in 3 parts: an improvement in track angle prediction (NN > Mom.), an improvement in the track error predictions (NN w/ weights > Mom. w/ weights) and an improvement from using a weighted scheme (NN w/ weights > NN, Mom. w/ weights > Mom. w/ cut).

Since (for an unbiased estimator) the FoM is proportional to $N_{\rm eff}^{-1/2} \sim t^{-1/2} $ , the FoM improvement corresponds to a 1-$1.2^{2} \approx 45$\% increase in effective exposure time {\it vs.} cut moments analysis, or $\approx 30$\% if moments weights are employed. The effective area peaks at $\sim 2.5$ keV, greatly favouring low energy tracks in the analyzed data set. Of course this disfavors our NN analysis which tends to perform best at high energies, but is conservative and realistic for soft astrophysical sources. Sources with harder spectra (e.g. highly absorbed AGN or accreting X-ray pulsars) will show even larger NN performance boosts. Note that we also do not include the events detected outside of the $2-8$\,keV energy range. A small MDP$_{99}$ improvement can be expected from inclusion of the many (poorly measured) low energy events, especially for very soft sources.

While all results shown here are for simulated tracks, we have applied this analysis to sample data from real GPD events, and find similar sensitivity improvements. More extensive comparisons with individual GPDs will be needed to optimize the analysis.

\begin{figure}
\centering
\includegraphics[width=1.0\textwidth]{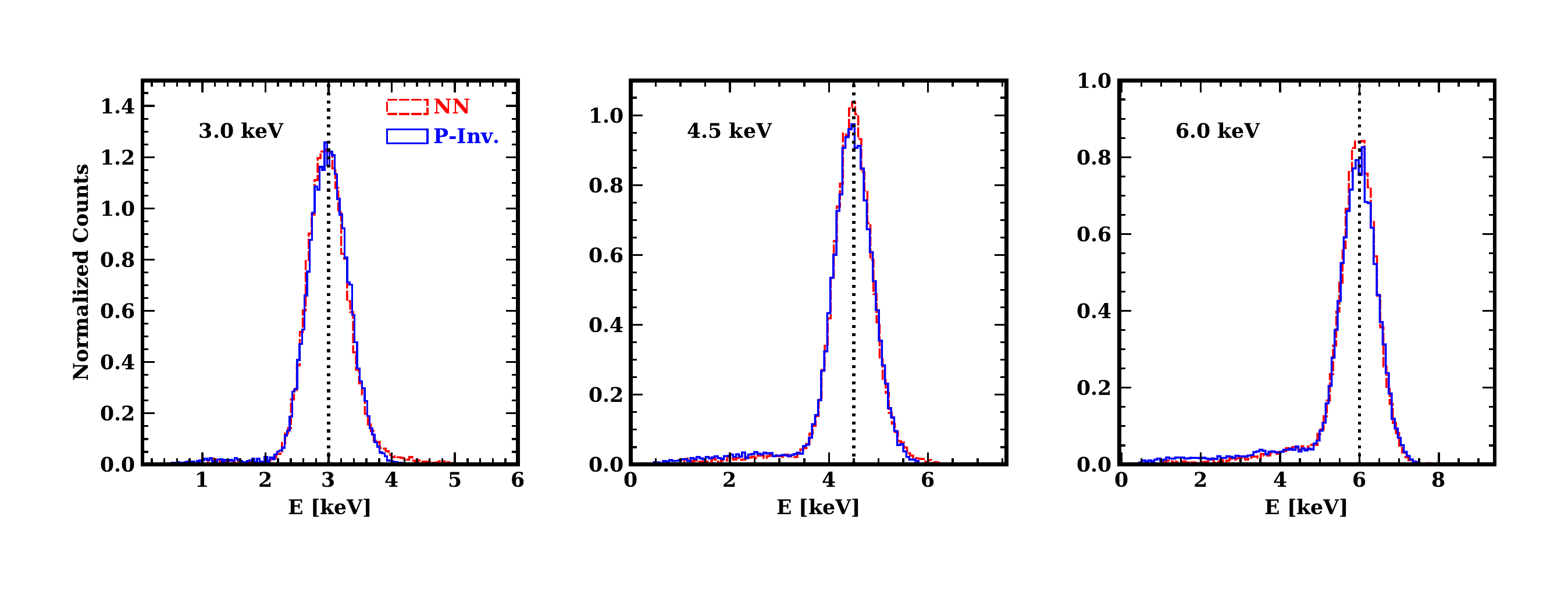}
\vskip -7mm
\caption{\textit{Left:} Recovered energy distributions. The main peak is slightly higher (smaller FWHM) for NN estimates, while the low energy tail of window/GEMS conversion partial tracks is somewhat suppressed.}
\label{fig:Eres1}
\end{figure} 

\subsection{Absorption points}

Our NN analysis also returns the photon absorption point and energy. Fig.~\ref{fig:example} shows the performance compared with the standard moment analysis for the localization in terms of root mean squared error (RMSE). In the standard pipeline, the localization shifts to the event barycenter (dashed curve) at low energies. It seems that our trained NN automatically shifts to a similar estimate at low energies as the red curve is better at all $E$. In practice the performance of all methods is adequate for IXPE imaging; the dotted line shows one quarter of the mirror half power diameter (HPD). However, the excellent NN photon localization on the detector can be very helpful in characterizing and mitigating position-dependent detector effects. We can compare our recovered absorption point localization HPDs at 2.7, 4.5, 6.4 and 8.0 keV (1.7, 1.6, 1.6, and 1.9 pixels, respectively) to those of \citet{li_electron_2017} (2.3, 3.4, 3.2, 3.7 pixel) and \citet{kitaguchi_convolutional_2019} (1.9, 1.6, 1.5, 1.6), finding that they compare favourably.

\subsection{Energy}

For the energy resolution, we find that the NN out-performs moments for all energies in the IXPE spectrum. Results are shown in figs.~\ref{fig:example} and \ref{fig:Eres1}.
While we expect that all sources observed by IXPE will have much higher resolution spectra available (e.g. from CCD data), this energy improvement can help in isolating polarization signatures to narrow spectra features, e.g. the Fe K$\alpha$ line. Further simulation with real astronomical spectra are needed to see if such line affects are within IXPE's reach.

\section{Summary and Discussion}

We have developed a state-of-the-art track reconstruction and polarization prediction framework for imaging X-ray polarimeters. We reconstruct X-ray photon track angles, absorption points and energies using deep ensembles; predictive errors in track angle are used to inform a weighted maximum likelihood estimation of source polarization parameters. We define a new FoM for polarization recovery, using bootstrap error distributions, which allows us to compare, on equal footing, our method with the IXPE project's current moment analysis, including cuts to the data sets and weights from the reconstruction quality measurements. We have tested our method with simulated and real detector data. On simple simulated power law spectra, our analysis implies that the increases to IXPE's polarization sensitivity provided by this method can increase the effective exposures by as much as $\sim 45\%$.  Preliminary results suggest improvements on real data are similar. 

Further verification with real X-ray calibration data sets is needed before this method can be applied to flight data. The networks would also need to be tuned, with calibration data, for the peculiarities of each flight detector. Of course the weighting scheme should be tuned for individual source spectra. Indeed, considering variation in Fig.\,\ref{fig:errors} the optimal weighting will be an energy dependent $\lambda(E_i)$. 
The method that we present here is general and may offer even larger gains for future missions with larger effective areas and higher spatial resolution from tighter mirror PSFs, such as eXTP and future proposed electron tracking polarization projects.

\section{Acknowledgements}

We would like to thank Bruce Tidor and Kevin Shi of MIT for early efforts related to this project and Marius Tirlea for discussions on statistical models. We would also like to thank the NASA FINESST grant 80NSSC19K1407 and NASA grant NNM17AA26C for supporting this work. The Italian contribution to IXPE is supported by the Italian Space Agency through the agreement ASI-INFN n.2017-13-H.O. Funding for this work was provided in part by contract 80MSFC17C0012 from the Marshall Space Flight Center (MSFC) to MIT in support of IXPE, a NASA Astrophysics Small Explorers mission.

\bibliographystyle{test}

\bibliography{references.bib}

\begin{thebibliography}{37}
\providecommand{\natexlab}[1]{#1}
\expandafter\ifx\csname urlstyle\endcsname\relax
  \providecommand{\doi}[1]{doi:\discretionary{}{}{}#1}\else
  \providecommand{\doi}{doi:\discretionary{}{}{}\begingroup
  \urlstyle{rm}\Url}\fi

\bibitem[{Krawczynski et~al.(2019)Krawczynski, Matt, Ingram, Taverna, Turolla,
  Kislat, Cheung, Bykov et~al.}]{krawczynski_using_2019}
H.~Krawczynski, G.~Matt, A.~R. Ingram, R.~Taverna, R.~Turolla, F.~Kislat,
  C.~C.~T. Cheung, A.~Bykov, et~al.
\newblock Using {X}-{Ray} {Polarimetry} to {Probe} the {Physics} of {Black}
  {Holes} and {Neutron} {Stars}.
\newblock 51:150, 2019.
\newblock Conference Name: Bulletin of the American Astronomical Society.

\bibitem[{Weisskopf et~al.(1976)Weisskopf, Cohen, Kestenbaum, Long, Novick, and
  Wolff}]{weisskopf_measurement_1976}
M.~C. Weisskopf, G.~G. Cohen, H.~L. Kestenbaum, K.~S. Long, R.~Novick, and
  R.~S. Wolff.
\newblock Measurement of the {X}-ray polarization of the {Crab} {Nebula}.
\newblock \emph{The Astrophysical Journal Letters}, 208:L125--L128, 1976.
\newblock ISSN 0004-637X.
\newblock \doi{10.1086/182247}.

\bibitem[{Costa et~al.(2001)Costa, Soffitta, Bellazzini, Brez, Lumb, and
  Spandre}]{costa_efficient_2001}
E.~Costa, P.~Soffitta, R.~Bellazzini, A.~Brez, N.~Lumb, and G.~Spandre.
\newblock An efficient photoelectric {X}-ray polarimeter for the study of black
  holes and neutron stars.
\newblock \emph{Nature}, 411:662--665, 2001.
\newblock ISSN 0028-0836.

\bibitem[{Bellazzini et~al.(2007)Bellazzini, Spandre, Minuti, Baldini, Brez,
  Latronico, Omodei, Razzano et~al.}]{bellazzini_sealed_2007}
R.~Bellazzini, G.~Spandre, M.~Minuti, L.~Baldini, A.~Brez, L.~Latronico,
  N.~Omodei, M.~Razzano, et~al.
\newblock A sealed {Gas} {Pixel} {Detector} for {X}-ray astronomy.
\newblock \emph{Nuclear Instruments and Methods in Physics Research A},
  579:853--858, 2007.
\newblock ISSN 0168-9002.
\newblock \doi{10.1016/j.nima.2007.05.304}.

\bibitem[{Feng and Bellazzini(2020)}]{feng_x-ray_2020}
H.~Feng and R.~Bellazzini.
\newblock The {X}-ray polarimetry window reopens.
\newblock \emph{Nature Astronomy}, 4(5):547--547, 2020.
\newblock ISSN 2397-3366.
\newblock \doi{10.1038/s41550-020-1103-6}.
\newblock Number: 5 Publisher: Nature Publishing Group.

\bibitem[{Sgrò and {IXPE Team}(2019)}]{sgro_imaging_2019}
C.~Sgrò and {IXPE Team}.
\newblock The {Imaging} {X}-ray {Polarimetry} {Explorer} ({IXPE}).
\newblock \emph{Nuclear Instruments and Methods in Physics Research A},
  936:212--215, 2019.
\newblock ISSN 0168-9002.
\newblock \doi{10.1016/j.nima.2018.10.111}.

\bibitem[{Zhang et~al.(2017)Zhang, Feroci, Santangelo, Dong, Feng, Lu, Nandra,
  Wang et~al.}]{zhang_extp_2017}
S.~N. Zhang, M.~Feroci, A.~Santangelo, Y.~W. Dong, H.~Feng, F.~J. Lu,
  K.~Nandra, Z.~S. Wang, et~al.
\newblock {eXTP}: {Enhanced} {X}-ray {Timing} and {Polarization} mission.
\newblock In \emph{Space {Telescopes} and {Instrumentation} 2016: {Ultraviolet}
  to {Gamma} {Ray}}, volume 9905, page 99051Q. International Society for Optics
  and Photonics, 2017.
\newblock \doi{10.1117/12.2232034}.

\bibitem[{Weisskopf(2018)}]{weisskopf_overview_2018}
M.~Weisskopf.
\newblock An {Overview} of {X}-{Ray} {Polarimetry} of {Astronomical} {Sources}.
\newblock \emph{Galaxies}, 6:33, 2018.
\newblock \doi{10.3390/galaxies6010033}.

\bibitem[{O'Dell et~al.(2018)O'Dell, Baldini, Bellazzini, Costa, Elsner, Kaspi,
  Kolodziejczak, Latronico et~al.}]{odell_imaging_2018}
S.~L. O'Dell, L.~Baldini, R.~Bellazzini, E.~Costa, R.~F. Elsner, V.~M. Kaspi,
  J.~J. Kolodziejczak, L.~Latronico, et~al.
\newblock The {Imaging} {X}-ray {Polarimetry} {Explorer} ({IXPE}): technical
  overview.
\newblock 0699:106991X, 2018.
\newblock \doi{10.1117/12.2314146}.
\newblock Conference Name: Space Telescopes and Instrumentation 2018:
  Ultraviolet to Gamma Ray.

\bibitem[{Muleri et~al.(2010)Muleri, Soffitta, Baldini, Bellazzini, Brez,
  Costa, Fabiani, Krummenacher et~al.}]{muleri_spectral_2010}
F.~Muleri, P.~Soffitta, L.~Baldini, R.~Bellazzini, A.~Brez, E.~Costa,
  S.~Fabiani, F.~Krummenacher, et~al.
\newblock Spectral and polarimetric characterization of the {Gas} {Pixel}
  {Detector} filled with dimethyl ether.
\newblock \emph{Nuclear Instruments and Methods in Physics Research A},
  620:285--293, 2010.
\newblock ISSN 0168-9002.
\newblock \doi{10.1016/j.nima.2010.03.006}.

\bibitem[{Bellazzini et~al.(2003)Bellazzini, Angelini, Baldini, Brez, Costa,
  Persio, Latronico, Massai et~al.}]{bellazzini_novel_2003}
R.~Bellazzini, F.~Angelini, L.~Baldini, A.~Brez, E.~Costa, G.~D. Persio,
  L.~Latronico, M.~M. Massai, et~al.
\newblock Novel gaseous x-ray polarimeter: data analysis and simulation.
\newblock In \emph{Polarimetry in {Astronomy}}, volume 4843, pages 383--393.
  International Society for Optics and Photonics, 2003.
\newblock \doi{10.1117/12.459381}.

\bibitem[{Moriakov et~al.(2020)Moriakov, Samudre, Negro, Gieseke, Otten, and
  Hendriks}]{moriakov_inferring_2020}
N.~Moriakov, A.~Samudre, M.~Negro, F.~Gieseke, S.~Otten, and L.~Hendriks.
\newblock Inferring astrophysical {X}-ray polarization with deep learning.
\newblock \emph{arXiv:2005.08126 [astro-ph]}, 2020.
\newblock ArXiv: 2005.08126.

\bibitem[{Graves et~al.(2006)Graves, Fernández, Gomez, and
  Schmidhuber}]{graves_connectionist_2006}
A.~Graves, S.~Fernández, F.~Gomez, and J.~Schmidhuber.
\newblock Connectionist temporal classification: labelling unsegmented sequence
  data with recurrent neural networks.
\newblock In \emph{Proceedings of the 23rd international conference on
  {Machine} learning}, {ICML} '06, pages 369--376. Association for Computing
  Machinery, Pittsburgh, Pennsylvania, USA, 2006.
\newblock ISBN 978-1-59593-383-6.
\newblock \doi{10.1145/1143844.1143891}.

\bibitem[{Young et~al.(2018)Young, Hazarika, Poria, and
  Cambria}]{young_recent_2018}
T.~Young, D.~Hazarika, S.~Poria, and E.~Cambria.
\newblock Recent {Trends} in {Deep} {Learning} {Based} {Natural} {Language}
  {Processing}.
\newblock \emph{arXiv:1708.02709 [cs]}, 2018.
\newblock ArXiv: 1708.02709.

\bibitem[{Tang et~al.(2019)Tang, Pan, Yin, and Khateeb}]{tang_recent_2019}
B.~Tang, Z.~Pan, K.~Yin, and A.~Khateeb.
\newblock Recent {Advances} of {Deep} {Learning} in {Bioinformatics} and
  {Computational} {Biology}.
\newblock \emph{Frontiers in Genetics}, 10, 2019.
\newblock ISSN 1664-8021.
\newblock \doi{10.3389/fgene.2019.00214}.
\newblock Publisher: Frontiers.

\bibitem[{Krizhevsky et~al.(2012)Krizhevsky, Sutskever, and
  Hinton}]{krizhevsky_imagenet_2012}
A.~Krizhevsky, I.~Sutskever, and G.~E. Hinton.
\newblock {ImageNet} {Classification} with {Deep} {Convolutional} {Neural}
  {Networks}.
\newblock In F.~Pereira, C.~J.~C. Burges, L.~Bottou, and K.~Q. Weinberger,
  editors, \emph{Advances in {Neural} {Information} {Processing} {Systems} 25},
  pages 1097--1105. Curran Associates, Inc., 2012.

\bibitem[{Brill et~al.(2019)Brill, Feng, Humensky, Kim, Nieto, and
  Miener}]{brill_investigating_2019}
A.~Brill, Q.~Feng, T.~B. Humensky, B.~Kim, D.~Nieto, and T.~Miener.
\newblock Investigating a {Deep} {Learning} {Method} to {Analyze} {Images} from
  {Multiple} {Gamma}-ray {Telescopes}.
\newblock \emph{2019 New York Scientific Data Summit (NYSDS)}, pages 1--4,
  2019.
\newblock \doi{10.1109/NYSDS.2019.8909697}.
\newblock ArXiv: 2001.03602.

\bibitem[{Choma et~al.(2018)Choma, Monti, Gerhardt, Palczewski, Ronaghi,
  Prabhat, Bhimji, Bronstein et~al.}]{choma_graph_2018}
N.~Choma, F.~Monti, L.~Gerhardt, T.~Palczewski, Z.~Ronaghi, Prabhat, W.~Bhimji,
  M.~M. Bronstein, et~al.
\newblock Graph {Neural} {Networks} for {IceCube} {Signal} {Classification}.
\newblock \emph{arXiv:1809.06166 [astro-ph, stat]}, 2018.
\newblock ArXiv: 1809.06166.

\bibitem[{Kitaguchi et~al.(2019)Kitaguchi, Black, Enoto, Hayato, Hill, Iwakiri,
  Kaaret, Mizuno, and Tamagawa}]{kitaguchi_convolutional_2019}
T.~Kitaguchi, K.~Black, T.~Enoto, A.~Hayato, J.~E. Hill, W.~B. Iwakiri,
  P.~Kaaret, T.~Mizuno, and T.~Tamagawa.
\newblock A convolutional neural network approach for reconstructing
  polarization information of photoelectric {X}-ray polarimeters.
\newblock \emph{Nuclear Instruments and Methods in Physics Research Section A:
  Accelerators, Spectrometers, Detectors and Associated Equipment}, 942:162389,
  2019.
\newblock ISSN 01689002.
\newblock \doi{10.1016/j.nima.2019.162389}.
\newblock ArXiv: 1907.06442.

\bibitem[{Tamagawa and {PRAXyS Team}(2017)}]{tamagawa_x-ray_2017}
T.~Tamagawa and {PRAXyS Team}.
\newblock X-ray {Polarimetry} {Mission} {PRAXyS}.
\newblock page 305, 2017.

\bibitem[{Lakshminarayanan et~al.(2017)Lakshminarayanan, Pritzel, and
  Blundell}]{lakshminarayanan_simple_2017}
B.~Lakshminarayanan, A.~Pritzel, and C.~Blundell.
\newblock Simple and scalable predictive uncertainty estimation using deep
  ensembles.
\newblock In \emph{Proceedings of the 31st {International} {Conference} on
  {Neural} {Information} {Processing} {Systems}}, {NIPS}'17, pages 6405--6416.
  Curran Associates Inc., Long Beach, California, USA, 2017.
\newblock ISBN 978-1-5108-6096-4.

\bibitem[{Efron and Tibshirani(1994)}]{efron_introduction_1994}
B.~Efron and R.~J. Tibshirani.
\newblock \emph{An {Introduction} to the {Bootstrap}}.
\newblock CRC Press, 1994.
\newblock ISBN 978-0-412-04231-7.
\newblock Google-Books-ID: gLlpIUxRntoC.

\bibitem[{Fort et~al.(2019)Fort, Hu, and Lakshminarayanan}]{fort_deep_2019}
S.~Fort, H.~Hu, and B.~Lakshminarayanan.
\newblock Deep {Ensembles}: {A} {Loss} {Landscape} {Perspective}.
\newblock \emph{arXiv:1912.02757 [cs, stat]}, 2019.
\newblock ArXiv: 1912.02757.

\bibitem[{Ovadia et~al.(2019)Ovadia, Fertig, Ren, Nado, Sculley, Nowozin,
  Dillon, Lakshminarayanan, and Snoek}]{ovadia_can_2019}
Y.~Ovadia, E.~Fertig, J.~Ren, Z.~Nado, D.~Sculley, S.~Nowozin, J.~V. Dillon,
  B.~Lakshminarayanan, and J.~Snoek.
\newblock Can {You} {Trust} {Your} {Model}'s {Uncertainty}? {Evaluating}
  {Predictive} {Uncertainty} {Under} {Dataset} {Shift}.
\newblock \emph{arXiv:1906.02530 [cs, stat]}, 2019.
\newblock ArXiv: 1906.02530.

\bibitem[{Kendall and Gal(2017)}]{kendall_what_2017}
A.~Kendall and Y.~Gal.
\newblock What {Uncertainties} {Do} {We} {Need} in {Bayesian} {Deep} {Learning}
  for {Computer} {Vision}?
\newblock In I.~Guyon, U.~V. Luxburg, S.~Bengio, H.~Wallach, R.~Fergus,
  S.~Vishwanathan, and R.~Garnett, editors, \emph{Advances in {Neural}
  {Information} {Processing} {Systems} 30}, pages 5574--5584. Curran
  Associates, Inc., 2017.

\bibitem[{Steppa and Holch(2019)}]{steppa_hexagdly_2019}
C.~Steppa and T.~L. Holch.
\newblock {HexagDLy} - {Processing} hexagonally sampled data with {CNNs} in
  {PyTorch}.
\newblock \emph{SoftwareX}, 9:193--198, 2019.
\newblock ISSN 23527110.
\newblock \doi{10.1016/j.softx.2019.02.010}.
\newblock ArXiv: 1903.01814.

\bibitem[{Agostinelli et~al.(2003)Agostinelli, Allison, Amako, Apostolakis,
  Araujo, Arce, Asai, Axen et~al.}]{agostinelli_geant4simulation_2003}
S.~Agostinelli, J.~Allison, K.~Amako, J.~Apostolakis, H.~Araujo, P.~Arce,
  M.~Asai, D.~Axen, et~al.
\newblock Geant4—a simulation toolkit.
\newblock \emph{Nuclear Instruments and Methods in Physics Research Section A:
  Accelerators, Spectrometers, Detectors and Associated Equipment},
  506(3):250--303, 2003.
\newblock ISSN 0168-9002.
\newblock \doi{10.1016/S0168-9002(03)01368-8}.

\bibitem[{He et~al.(2015)He, Zhang, Ren, and Sun}]{he_deep_2015}
K.~He, X.~Zhang, S.~Ren, and J.~Sun.
\newblock Deep {Residual} {Learning} for {Image} {Recognition}.
\newblock \emph{arXiv:1512.03385 [cs]}, 2015.
\newblock ArXiv: 1512.03385.

\bibitem[{Huang et~al.(2018)Huang, Liu, van~der Maaten, and
  Weinberger}]{huang_densely_2018}
G.~Huang, Z.~Liu, L.~van~der Maaten, and K.~Q. Weinberger.
\newblock Densely {Connected} {Convolutional} {Networks}.
\newblock \emph{arXiv:1608.06993 [cs]}, 2018.
\newblock ArXiv: 1608.06993.

\bibitem[{Sutskever et~al.(2013)Sutskever, Martens, Dahl, and
  Hinton}]{sutskever_importance_2013}
I.~Sutskever, J.~Martens, G.~Dahl, and G.~Hinton.
\newblock On the importance of initialization and momentum in deep learning.
\newblock In \emph{Proceedings of the 30th {International} {Conference} on
  {International} {Conference} on {Machine} {Learning} - {Volume} 28},
  {ICML}'13, pages III--1139--III--1147. JMLR.org, Atlanta, GA, USA, 2013.

\bibitem[{Kislat et~al.(2015)Kislat, Clark, Beilicke, and
  Krawczynski}]{kislat_analyzing_2015}
F.~Kislat, B.~Clark, M.~Beilicke, and H.~Krawczynski.
\newblock Analyzing the data from {X}-ray polarimeters with {Stokes}
  parameters.
\newblock \emph{Astroparticle Physics}, 68:45--51, 2015.
\newblock ISSN 0927-6505.
\newblock \doi{10.1016/j.astropartphys.2015.02.007}.

\bibitem[{Karampatziakis and Langford(2011)}]{karampatziakis_online_2011}
N.~Karampatziakis and J.~Langford.
\newblock Online importance weight aware updates.
\newblock In \emph{Proceedings of the {Twenty}-{Seventh} {Conference} on
  {Uncertainty} in {Artificial} {Intelligence}}, {UAI}'11, pages 392--399. AUAI
  Press, Barcelona, Spain, 2011.
\newblock ISBN 978-0-9749039-7-2.

\bibitem[{Hu and Zidek(2002)}]{hu_weighted_2002}
F.~Hu and J.~V. Zidek.
\newblock The weighted likelihood.
\newblock \emph{Canadian Journal of Statistics}, 30(3):347--371, 2002.
\newblock ISSN 1708-945X.
\newblock \doi{10.2307/3316141}.

\bibitem[{Wächter and Biegler(2006)}]{wachter_implementation_2006}
A.~Wächter and L.~T. Biegler.
\newblock On the implementation of an interior-point filter line-search
  algorithm for large-scale nonlinear programming.
\newblock \emph{Mathematical Programming}, 106(1):25--57, 2006.
\newblock ISSN 1436-4646.
\newblock \doi{10.1007/s10107-004-0559-y}.

\bibitem[{Weisskopf et~al.(2010)Weisskopf, Elsner, and
  O'Dell}]{weisskopf_understanding_2010}
M.~C. Weisskopf, R.~F. Elsner, and S.~L. O'Dell.
\newblock On understanding the figures of merit for detection and measurement
  of x-ray polarization.
\newblock \emph{arXiv:1006.3711 [astro-ph]}, page 77320E, 2010.
\newblock \doi{10.1117/12.857357}.
\newblock ArXiv: 1006.3711.

\bibitem[{Weisskopf et~al.(2016)Weisskopf, Ramsey, O’Dell, Tennant, Elsner,
  Soffita, Bellazzini, Costa et~al.}]{weisskopf_imaging_2016}
M.~C. Weisskopf, B.~Ramsey, S.~L. O’Dell, A.~Tennant, R.~Elsner, P.~Soffita,
  R.~Bellazzini, E.~Costa, et~al.
\newblock The {Imaging} {X}-ray {Polarimetry} {Explorer} ({IXPE}).
\newblock \emph{Results Phys}, 6:1179--1180, 2016.
\newblock ISSN 2211-3797.
\newblock \doi{10.1016/j.rinp.2016.10.021}.

\bibitem[{Li et~al.(2017)Li, Zeng, Feng, Cang, Li, Zhang, Zeng, Cheng
  et~al.}]{li_electron_2017}
T.~Li, M.~Zeng, H.~Feng, J.~Cang, H.~Li, H.~Zhang, Z.~Zeng, J.~Cheng, et~al.
\newblock Electron {Track} {Reconstruction} and {Improved} {Modulation} for
  {Photoelectric} {X}-ray {Polarimetry}.
\newblock \emph{Nuclear Instruments and Methods in Physics Research Section A:
  Accelerators, Spectrometers, Detectors and Associated Equipment}, 858:62--68,
  2017.
\newblock ISSN 01689002.
\newblock \doi{10.1016/j.nima.2017.03.050}.
\newblock ArXiv: 1611.07244.

\end{thebibliography}

\end{document}